\documentclass[12pt]{article}
\usepackage{latexsym,amsmath,amssymb}

\newtheorem{theorem}{Theorem}[section]

\newtheorem{lemma}[theorem]{Lemma}
\newtheorem{proposition}[theorem]{Proposition}
\newtheorem{corollary}[theorem]{Corollary}

\newtheorem{definition}[theorem]{Definition}

\newtheorem{remark}[theorem]{Remark}
\newtheorem{question}[theorem]{Question}

\newtheorem{convention_ital}[theorem]{Convention} 

\newcommand{\jpcom}[1]{}
\newcommand{\jpcor}[1]{}

\newenvironment{convention}{\begin{convention_ital}\rm }{\end{convention_ital}}

\newcommand{\asle}{\prec}
\newcommand{\asge}{\succ}

\newcommand{\ignore}[1]{}
\newcommand{\widebar}{\overline}
\newcommand{\wb}{\widebar}
\newcommand{\wt}{\widetilde}

\newcommand{\proof}{\par\noindent {\bf Proof}\space\space}
\newcommand{\proofbox}{\begin{flushright}$\Box$\end{flushright}}

\newcommand{\from}{\colon}
\newcommand{\supp}{{\rm Supp}}
\newcommand{\finite}{{\rm fn}}
\newcommand{\Dirichlet}{{\rm Dir}}
\newcommand{\Lipschitz}{{\rm Lip}}

\newcommand{\nablacalc}{{\nabla_{\rm calc}}}

\newcommand{\Z}{{\bf Z}}      
\newcommand{\R}{{\bf R}}  
\newcommand{\reals}{\R}

\newcommand{\integers}{\Z}

\newcommand{\vtr}{{\cal V}_{\rm T}}
\newcommand{\cg}{{\cal G}}
\newcommand{\tcg}{{\rm T}{\cal G}}
\newcommand{\ch}{{\cal H}}
\newcommand{\dv}{d{\cal V}}
\newcommand{\de}{d{\cal E}}
\newcommand{\cv}{{\cal V}}
\newcommand{\ce}{{\cal E}}
\newcommand{\etr}{{\cal E}_{\rm T}}
\newcommand{\cf}{{\cal F}}

\newcommand{\cd}{{\cal D}}
\newcommand{\area}{{\cal A}}
\newcommand{\cq}{{\cal Q}}
\newcommand{\cm}{{\cal M}}
\newcommand{\cl}{{\cal L}}

\newcommand{\inte}[1]{{\int #1\,\de}}
\newcommand{\intv}[1]{{\int #1\,\dv}}

\newcommand{\closure}[1]{{\widebar{#1}}}
\renewcommand{\complement}[1]{{#1^{\rm c}}}
\newcommand{\interior}[1]{{#1^{\rm o}}}

\newcommand{\ex}[2]{\{#1\}^{#2}}

\newcommand{\nor}{{\bf n}}

\title{Calculus on Graphs}

\author{Joel Friedman\thanks{Research supported in part by an NSERC grant.}
	 \\ Department of Mathematics \\ 
	University of British Colubmia \\ Vancouver, BC  V6T 1Z2 \\ CANADA
	\\ {\tt jf@@math.ubc.ca} 
\and
	Jean-Pierre Tillich\thanks{Research supported in part
by a NSERC Postdoctoral Fellowship. This author enjoyed the
hospitality
of the University of British Colombia (Vancouver, Canada) while part
of this research was carried out. }
	 \\ LRI b\^atiment 490 \\ Universit\'e Paris-Sud \\ Orsay 91405 \\
	FRANCE \\ {\tt tillich@@lri.fr}}

\begin{document}           
\maketitle                 

\section{Introduction}

The purpose of this paper is to develop a ``calculus'' on graphs that
allows graph theory to have new connections to analysis.  
For example, our framework gives rise to many new partial differential
equations on graphs, most notably a new (Laplacian based) wave equation
(see \cite{friedman_tillich_distance,friedman_tillich_wave}); 
this wave equation gives rise to a
partial improvement on the Chung-Faber-Manteuffel diameter/eigenvalue
bound in graph theory
(see \cite{chung_faber_manteuffel}), and the
Chung-Grigoryan-Yau and (in a certain case) Bobkov-Ledoux distance/eigenvalue
bounds in analysis 
(see \cite{chung_grigoryan_yau_1,chung_grigoryan_yau_2,bobkov_ledoux}).
Our framework also allows most techniques for the non-linear $p$-Laplacian
in analysis to be easily carried over to graph theory
(see \cite{friedman_tillich_plaplacian}).

After developing the core ``calculus on graphs'' common to this and future
works, we give some new variants and simpler proofs of inequalities, such
as those of 
Federer-Fleming and Sobolev, known in graph theory
(as in \cite{MR97d:60114,MR99b:60119,MR94d:46075,MR97a:46040,
MR97c:46039,MR98e:58159,MR97j:47055,ChuYausob}).  We also develop a 
notion of ``split'' functions
that gives some improvements in these
inequalities in certain cases in both analysis and graph theory.
Yet, it might be said that the applications of ``calculus on graphs''
given here are secondary to the future applications to appear in
\cite{friedman_tillich_distance,friedman_tillich_wave,
friedman_tillich_plaplacian}.

One key point in this ``calculus on graphs'' is that, for what appears
to be the first time, ``non-linear'' functions
(functions that are not edgewise linear) become important; in previous
approaches
that unify graph theory and analysis (e.g. \cite{Fried2} and the references
there) only linear functions are ultimately used.  The use of
non-linear functions allows many proofs and ideas to carry over more simply
from analysis to graphs and vice versa.

We caution the reader that using ``non-linear'' techniques can lead
to a small loss in the resulting constants in the inequalities.  However,
it is almost always the case that the most interesting aspects of the
inequalities (their
dependencies on geometric constants, critical exponents,
the statements of interesting theorems based on the inequalities,
etc.) carry over identically.  Also, when there is a loss in the constant
(and this isn't always the case), one can usually easily recover this loss by
slightly refining the analysis done with the simpler, non-linear
technique.

Another benefit of the calculus on graphs is that it enables more analysis
techniques to carry over to graphs and vice versa in a very direct and
simple fashion; less intuition is obscured in technicalities that are
particular to analysis or graphs.

We mention that most of the inequalities we prove in this paper can be
called {\em gradient inequalities}, in which we lower bound the $L^p$ of
the gradient of $f$ in terms of norms on $f$ and constants depending on
the graph (``isoperimetric constants'').  A large number of well known
results in graph theory such as results on the eigenvalues of the Laplacian
can be viewed as gradient inequalities. 

In section 2 we develop some notions of ``calculus on graphs.''
In section 3 we discuss some preliminary remarks on gradient inequalities
for ``closed'' graphs, i.e. finite graphs with no boundary; gradient
inequalities for such graphs are only interesting when one works with 
functions ``modulo constant functions.''
In section 4 we discuss Federer-Fleming theorems; these give lower bounds
of the $L^1$ norm of the gradient.  In section 5 we discuss analogues of
Cheeger's inequality for graphs, namely the inequalities of Dodziuk, Alon,
and Mohar, from our point of view.
In section 6 we discuss the heat kernel, in preparation for section 7.
In section 7 we discuss lower bounds for the $L^p$ norm of
the gradient, including Sobolev inequalities and Nash inequalities and
the resulting eigenvalue inequalities.
In section 7 our techniques, especially using that of ``split'' functions
in the closed case, improves the constants in
results published previously; 
some 
other material in sections 4-7 represents generalizations, alternate forms,
and/or 
simplifications
of theorems that appears in the literature
(see, for example, \cite{MR97d:60114,MR99b:60119,MR94d:46075,MR97a:46040,
MR97c:46039,MR98e:58159,MR97j:47055,ChuYausob}).

\noindent
{\bf Notation}\\
Throughout this paper, if $1\le p\le\infty$, then $p'$ is the dual exponent
of $p$, meaning that $1\le p'\le\infty$ with $(1/p)+(1/p')=1$; then
$(p-1)(p'-1)=1$. We let $f^+= \max(f,0)$ and $f^-=\max(-f,0)$.

If $\Omega$ is a subset of a topological space, then $\closure{\Omega}$
denotes $\Omega$'s closure, and $\complement{\Omega}$ denotes $\Omega$'s
complement.

If $G=(V,E)$ is a graph then $e\sim\{u,v\}$ means that $e$ is an edge whose
endpoints are $u$ and $v$ (since we will allow multiple edges, we cannot
replace the $\sim$ with an $=$).  We write $v\in e$ or $e\ni v$ to
mean that $e$ has $v$ as an endpoint.  We write $u\sim v$ to indicate
that $u$ and $v$ have an edge joining them.

\section{Calculus on Graphs}

\subsection{The Setup}
We use a similar setting as in \cite{Fried2}, and we recall this setting
here.
Let $G=(V,E)$ be a graph (undirected), such that with each edge, 
$e\in E$, we have
associated a length, $\ell_e>0$.
We form the {\em geometric realization}, $\cg$, of $G$, which is the
metric space consisting of $V$ and a closed
interval of length $\ell_e$ from $u$ to
$v$ for each edge $e=\{u,v\}$.  
When there is no confusion,
we identify a $v\in V$ with its corresponding point in $\cg$ and identify
an $e\in E$ with its corresponding closed interval in $\cg$.  By an
{\em edge interior} we mean the interior of an edge in $\cg$.

In analysis, a Riemannian manifold may or may not have a {\em boundary}.
However, certain concepts, such as {\em nodal regions} and certain eigenvalue
inequalities, require the notion of a boundary {\em even if the original
manifold has no boundary}.  The graph theoretic analogues of these concepts
will also require the notion of a boundary in the graph setting.  

\begin{definition} The {\em boundary}, $\partial\cg$, of a graph, $\cg$,
is simply a specified subset of its vertices.  By the {\em interior of
$\cg$}, denoted $\interior\cg$, we mean $\cg\setminus\partial\cg$; similarly
the {\em interior vertices}, denoted $\interior V$, we mean $V\setminus
\partial\cg$.
\end{definition}

\begin{convention} By a {\em traditional graph} we mean an undirected graph
$G=(V,E)$.  Throughout this article we assume our graphs are always
given with (1) lengths associated to each edge, (2) a specified boundary
(i.e. a specified subset of vertices).  Whenever an edge length is not
specified, it is taken to be one.  Whenever a boundary is not specified,
it is taken to be empty.  We refer to the geometric realization, $\cg$,
of the graph
as the graph, when no confusion may arise.
\end{convention}

\begin{definition} By $C^k(\cg)$ (respectively $C^k(\cg\setminus V)$), 
the set of {\em $k$-times continuously
differentiable functions on $\cg$ (respectively $\cg\setminus V$)} we 
mean the set of 
continuous
functions on $\cg$ (respectively $\cg\setminus V$) 
whose restriction to each edge interior is $k$-times
uniformly continuously differentiable (as a function on that real interval).
\end{definition}

We cannot differentiate functions on $\cg$ without orienting the edges;
however, we can always take the gradient of a differentiable function
as long as we know what is meant by a vector field.  Recall that a vector
field on an interval is a section of its tangent bundle or, what is the same,
a function on the interval with an orientation of the interval, where we
identify $f$ plus an orientation with $-f$ with the opposite orientation.

\begin{definition} By $C^k(\tcg)$, the set of $k$-times continuously 
differentiable vector fields on $\cg$, we mean those data consisting of a
$k$-times uniformly
continuously differentiable vector field on each open interval
corresponding to each edge interior.
\end{definition}

Notice that a vector field is not defined on at a vertex, rather only on edge
interiors.

\begin{definition} For $f\in C^k(\cg)$ we may form, by differentiation,
its {\em gradient}, $\nabla f\in C^{k-1}(\tcg)$.  For $X\in C^k(\tcg)$
we can form, by differentiation, its {\em calculus divergence},
$\nablacalc\cdot X\in C^{k-1}(\tcg\setminus V)$.
\end{definition}

Many theorems in analysis apply only to smooth functions or vector fields
of {\em compact support}.  Similarly, here, in working with infinite
graphs (i.e. when either $V$ or $E$ or both are infinite),
our theorems will only apply to a smaller class than $C^k$.

\begin{definition} A subset $\Omega\subset\cg$ is of {\em finite type}
if it lies in the union of finitely many vertices and edges.  A function
on $\cg$ is of {\em finite type} if its support
(i.e. the closure of the set where it does
not vanish) is of finite type.
We set
$C^k_\finite(\cg)$ to be those elements
of $C^k(\cg)$ of finite type;
we similarly define
$C^k_\finite(\cg\setminus V)$ and $C^k_\finite(\tcg)$.
\end{definition}
Notice that for a finite graph, i.e. when $E$ and $V$ are finite, every
set is of finite type and $C^k_\finite$ coincides with $C^k$.  
Notice also that in general a set of finite type is relatively compact
in the metric space topology on $\cg$ (i.e. its closure is compact),
but not conversely--- indeed, if $\cg$ consists of a vertex and a self-loop
(i.e. edge from $v$ to $v$)
of length $1/n$ for each integer $n$, then $\cg$ itself is compact but
not of finite type.

Another further subclass of $C^k(\cg)$ will be very important.

\begin{definition} An $f\in C^k_\finite(\cg)$ is said to satisfy the
{\em Dirichlet condition} if $f$ vanishes on $\partial\cg$.  We let
$C^k_\Dirichlet(\cg)$ denote the set of such functions.
\end{definition}
If $\partial\cg$ is empty then clearly $C^k_\Dirichlet(\cg)=C^k_\finite(\cg)$.

Finally, the positive or negative part of a smooth function will usually
only be Lipschitz continuous.  We therefore need the following definition.

\begin{definition} $\Lipschitz(\cg)$ denotes the class of {\em Lipschitz
continuous} functions on $\cg$, i.e. those $f\in C^0(\cg)$ whose restriction
to each edge interior is uniformly Lipschitz continuous.  We similarly
define $\Lipschitz_\finite(\cg)$ and $\Lipschitz_\Dirichlet(\cg)$.
\end{definition}

\subsection{Two Volume Measures}

In analysis concepts such as Laplacians, Rayleigh quotients, and isoperimetric
constants are defined using one volume measure; in calculus on graphs we use
two ``volume'' measures.

\begin{definition} A {\em vertex measure}, $\cv$, is a measure
supported on $V$ with $\cv(v)>0$ for all $v\in V$.  An {\em edge measure}, 
$\ce$, is a measure with $\ce(v)=0$ for all $v\in V$ and whose restriction
to any edge interior, $e\in E$, is Lebesgue measure 
(viewing the interior as an open interval) times a constant $a_e>0$.
\end{definition}

Traditional graph theory usually works with the traditional vertex and edge
measures, $\vtr$ and $\etr$, given by $\vtr(v)=1$ for all $v\in V$ and
$a_e=1$ for all $e\in E$, i.e. $\etr$ is just Lebesgue measure at each edge.

\begin{convention} Henceforth we assume that any graph has associated with
it a vertex measure, $\cv$, and an edge measure, $\ce$.  When $\cv$ is not
specified we take it to be $\vtr$; similarly, when unspecified
we take $\ce$ to be $\etr$.
\end{convention}

In this article we write
$$
\inte{f} \qquad\mbox{and}\qquad \intv{f}
$$
for
$$
\int_{\cg} f \,d\ce \qquad\mbox{and}\qquad \int_{\cg} f \,d\cv.
$$

In this article, if $\mu$ is a measure on $\cg$,
$1\le p\le \infty$, and $f$ a is real or vector-valued
on $\cg$ with $|f|$ $\mu$-measurable, then we define the usual $L^p(\cg,\mu)$
norm
$$
\|f\|_{p,\mu} = \left( \int_\cg |f|^p\,d\mu \right)^{1/p} 
$$
(with $p=\infty$ we take the norm to be the essential supremum of $|f|$
with respect to $\mu$).

\begin{convention} If $f\from\cg\to\reals$ is measurable, then $\|f\|_p$
means $\|f\|_{p,\cv}$ unless otherwise mentioned.  For 
$X\in C^0(\tcg)$, $\|X\|_p$ means $\|X\|_{p,\ce}$.
\end{convention}

\subsection{Three Rayleigh Quotients}

In this subsection we pause to give an example of translating results
from analysis to results in graph theory.  By a {\em Rayleigh quotient} we
mean a functional
$$
{\cal R}(f) = \frac{\int{ |\nabla f|^2}\,d\mu_1}{\int{ |f|^2\,d\mu_2}}
$$
defined on some class of functions, $f$, in a setting where the above quotient
has some reasonable interpretation.  We shall now give three precise settings
where this Rayleigh quotient makes sense.

\begin{enumerate}
\item {\bf Analysis or Riemannian geometry:} 
the above Rayleigh quotient appears
with $d\mu_1=d\mu_2=dV_g$, the Riemannian volume (or the usual volume in
$\reals^n$):
$$
{\cal R}(f) = \frac{\int |\nabla f(x)|^2 \,dV_g(x)}{\int f^2(x)\,dV_g(x)} .
$$
\item {\bf Traditional graph theory:}
here a function is a function on the vertices of the graph.  To define
$\nabla f$ we (arbitrarily and unnaturally) fix an orientation for
each edge; we declare $\nabla f$ on an
edge oriented $(u,v)$ to be $f(v)-f(u)$; the gradient is therefore a real
number.  The measures $d\mu_i$, $i=1,2$ are taken to be the edge counting
and vertex counting measures.  The Rayleigh quotient becomes:
$$
{\cal R}(f) = 
\frac{\sum_{\{u,v\}\in E}\bigl( f(u)-f(v) \bigr)^2}{\sum_{v\in V} f^2(v)}
$$
In this setting notions from analysis
such as those of nodal regions, level sets of a function, etc.
do not have an exact translation; proofs of theorems in traditional
graph theory that implicitly
use such notions may be unnecessarily awkward.
\item {\bf Calculus on graphs:}
here a function is a function on the geometric realization.  $\nabla f$ is
defined as above, and $d\mu_1=d\ce$, $d\mu_2=d\cv$, and the Rayleigh quotient
is
$$
{\cal R}(f) = \frac{\inte{ |\nabla f|^2}}{\intv{ |f|^2}} .
$$
Many more concepts and theorems in analysis translate almost immediately to
this setting.  It is usually easy to see that theorems in graph theory are
the same whether one states them in this setting or in traditional graph
theory.  

For example, consider a minimizer, $f$, of ${\cal R}(f)$ subject
to certain conditions on $f$'s values at the vertices.  It is easy to see that
such a minimizer must be {\em edgewise linear}, i.e. linear
when restricted to any edge interior
(see proposition~\ref{pr:linear}); hence
by restricting $f$ to its values at the vertices and taking the Rayleigh
quotient of traditional graph theory we get the same Rayleigh quotient as 
here.
\end{enumerate}

Consider a Rayleigh quotient in traditional graph theory that we wish obtain
in calculus on graphs when restricted to edgewise linear functions.
Then the $a_e/\ell_e$'s
and $\cv(v)$'s are determined up to a multiplicative constant, since
$$
\intv{ |f|^2}=\sum_v |f^2(v)|\cv(v) \quad\mbox{and}\quad
\inte{ |\nabla f|^2} = \sum_{e=\{u,v\} \in E} |f(u)-f(v)|^2 (a_e/\ell_e)
$$
for edgewise linear $f$.
However, we have the freedom to set either $a_e$ or $\ell_e$ as we please;
at times there are reasons to set one or the other to $1$.

\subsection{Half-degree and simple inequalities}

If $f$ is {\em edgewise linear}, i.e. $f$ is continuous and its restriction
to each edge interior is a linear function, 
then clearly
$$
\int_e f\,\de = \Bigl( f(u)+f(v)\Bigr) \ce(e)/2
$$
for each edge $e=\{u,v\}$.  Hence
\begin{equation}\label{eq:derhodv}
\inte{f} = \intv{f\rho},
\end{equation}
where 
\begin{equation}\label{eq:rho}
\rho(v)=\cv(v)^{-1} \sum_{e\ni v} \ce(e)/2.
\end{equation}
\begin{definition} The {\em half-degree} of a vertex, $v$, is $\rho(v)$ as
defined above.  We denote the infimum and supremum of $\rho$'s values by
$\rho_{\inf}$ and $\rho_{\sup}$.  
\end{definition}
In traditional graph theory, where $\ce=\etr$ and $\cv=\vtr$ we have that
$\rho(v)$ is just one-half the degree of $v$.

By convention a graph, $\cg$, encompasses the specification of $\cv$ and $\ce$;
hence we may write $\rho_{\sup}(\cg)$ and $\rho_{\inf}(\cg)$ without ambiguity.
\begin{definition} We say that a graph $\cg$ is {\em $r$-regular} if 
$\rho_{\inf}(\cg)=\rho_{\sup}(\cg)=r/2$.
\end{definition}

Clearly we have:
\begin{proposition} Let $f\in C^0_\finite(\cg)$ be an
{\em edgewise convex} function,
i.e. its restriction to each edge is convex\footnote{
We mean $f\bigl( \alpha x + (1-\alpha)y\bigr) \le \alpha f(x) + (1-\alpha)
f(y)$ for all $\alpha\in[0,1]$ and $x,y$ on the open edge interval.  So
$f$ does not have to be $C^2$ or $C^1$; 
e.g. the function $|x-1/2|$ is convex on
$[0,1]$.
}
(not necessary strictly convex); further assume that $f$ is non-negative
on all vertices.
Then
\begin{equation}\label{eq:convex}
\inte{f} \le \rho_{\sup} \intv{f}.
\end{equation}
Similarly for a non-negative edgewise concave function we have
\begin{equation}\label{eq:concave}
\inte{f} \ge \rho_{\inf} \intv{f}.
\end{equation}
If $\cg$ is regular then we can drop the requirement that $f$ be non-negative
at the vertices.
\end{proposition}
\proof We will show equation~\ref{eq:convex}; equation~\ref{eq:concave}
follows similarly.

Let $\wt f$ be the edgewise linear function whose values at the vertices
agree with those of $f$.  Then $\wt f\ge f$ and so $\inte{f}\le\inte{\wt f}$.
The non-negativity of $\wt f$ at the vertices implies that
$$
\intv{\rho\wt f} \le \rho_{\sup}\intv{\wt f}.
$$
These inequalities and equation~\ref{eq:derhodv} imply
$$
\inte{f}\le\inte{\wt f}=\intv{\rho\wt f}\le \rho_{\sup}\intv{\wt f}
=\rho_{\sup}\intv{f},
$$
using the fact that $f$ and $\wt f$ agree on the vertices.
We conclude equation~\ref{eq:convex}.  Furthermore, if $\cg$ is regular,
then $\intv{\rho\wt f}=\rho_{\sup}\intv{\wt f}$ for any $\wt f$, and we can
drop the
requirement that $f$'s values be non-negative at the vertices.
\proofbox

The above proposition has one special case that we will often need in 
establishing inequalities.  
\begin{corollary}
For any edgewise linear $f$ we have
\begin{equation}\label{eq:rhoconcaveineq}
\| f \|_{p,\ce} \le \rho_{\sup}^{1/p} \| f \|_{p,\cv}
\end{equation}
for any $p\ge 1$.
\end{corollary}
\proof It suffices to show that
$$
\inte{|f|^p}\le \rho_{\sup}\intv{|f|^p}.
$$
But $|f|^p$ is edgewise convex for $p\ge 1$.
\proofbox

Note that in analysis one has $\cv$ and $\ce$ replaced by the same measure,
and so the two integrals above are the same.  The closest we can seem to
come to this
in graph theory is
$$
\inte{f} = \intv{f}
$$
for $2$-regular graphs and edgewise
linear functions, $f$.  Of course, a $d$-regular
graph becomes a $2$-regular graph if either $\ce$ or $\cv$ is scaled 
appropriately.

We remark that given $\ce$ such that for any vertex, $v$, we have
$$
\sum_{e\ni v} \ce(e) \;<\infty,
$$
there is a unique measure $\cv$ such that $\cg$ is $2$-regular.  We call
$\cv$ the {\em natural} measure with respect to $\ce$.

We finish by discussing $\rho$ in other settings.
\begin{enumerate}
\item
If $G$ is viewed as an irreducible, reversible Markov chain, with stationary
distribution $\pi$ and transition probabilities $K$, then for the typical
Rayleigh quotient used we have
$\cv_{\rm P}(u)=\pi(u)$ and $a_e/\ell_e=\pi(u)K(u,v)=
\pi(v)K(v,u)$.  If we insist upon $\ell_e=1$,
then we have $\ce_{\rm P}(e)=a_e\ell_e=a_e=\pi(u)K(u,v)=\pi(v)K(v,u)$.
Since $\sum_v K(u,v)=1$ for fixed $u$, we have that
$(\cv_{\rm P},\ce_{\rm P})$ is $1$-regular.
\item
In \cite{dodziuk-kendall,dodziuk-karp,chung},
the denominator of the traditional Rayleigh quotient is
modified from the sum of $f^2(v)$ to that of
$d_v f^2(v)$, where $d_v$ is the degree
of $v$.  This corresponds to taking $\ce=\etr$ and $\cv(v)=d_v$.
In this case $\rho(v)=1/2$ for all $v$, i.e. $\cg$ is $1$-regular.
\end{enumerate}

\subsection{Remarks on Square Norm Inequalities}

In this subsection we make two remarks on the inequality in 
equation~\ref{eq:rhoconcaveineq} for $p=2$, namely
\begin{equation}\label{eq:squarenorm}
\| f \|_{2,\ce} \le \rho_{\sup}^{1/2} \| f \|_{2,\cv}
\end{equation}
This inequality comes up a lot in Laplacian eigenvalues and therefore
Cheeger's inequality.

Our first remark is that while the analogue of this inequality cannot be
improved upon in analysis (for there $\ce=\cv$), it can be improved, in
a certain sense, in graph theory.  Namely, if $f$ is linear on an edge
$e=\{u,v\}$ and $f(u)=b$ and $f(v)=c$, then it is easy to see that
$$
\int_e f^2\,d\ce = a_e(b^2+bc+c^2)/3,\quad\mbox{and}\quad
\int_e |\nabla f|^2\,d\ce = a_e(b^2-2bc+c^2),
$$
provided that all edge lengths are $1$.  We easily conclude that
\begin{equation}\label{eq:mohar}
\|f\|_{2,\ce}^2 + (1/6)\|\nabla f\|_2^2 = \intv{\rho f^2}\le 
\rho_{\sup}\|f\|_{2,\cv}^2,
\end{equation}
with equality if $\cg$ is regular.  This improvement to
equation~\ref{eq:squarenorm} is interesting in Laplacian eigenvalues and
Cheeger's inequality, for there it is precisely $\|\nabla f\|_2$ that one
is interested in and bounding from below.  This type of improvement
seems to have been first
exploited by Mohar (see \cite{mohar-inequalities,mohar-numbers}).

Our second remark is that later in this article we will be interested
in upper bounding $\|fX\|_{2,\ce}/\|f\|_{2,\cv}$ for a function $f$ and
an edgewise constant vector field, $X$.  To do so we simply remark that
\begin{equation}\label{eq:needed_for_alon}
\|fX\|_{2,\ce}\le \rho_{\sup}^{1/2}(\cg^{|X|^2})\|f\|_{2,\cv},
\end{equation}
where $\cg^{|X|^2}$ is the graph obtained from $\cg$ by multiplying the
edge weights, $a_e$, by $|X(e)|^2$.

\subsection{The Divergence}

The divergence of a vector field and the Laplacian of a function
can be defined in terms of concepts that
are already fixed, namely a graph (encompassing measures $\ce$ and $\cv$)
and the gradient.  Interestingly enough, the divergence turns out to be
different from the ``calculus divergence'' described earlier.

Before defining the divergence we record a ``divergence theorem'' for the
calculus divergence.

Let $X\in C^1(\tcg)$.
For any edge $e=\{u,v\}$ let $X|_e$ denote $X$ restricted to the interior of
$e$ and then extended to $u$ and $v$ by continuity.  We clearly have
$$
\int_e \nablacalc\cdot X \,\de = a_e \Bigl( \nor_{e,u}\cdot X|_e(u) + 
\nor_{e,v} \cdot X|_e(v) \Bigr),
$$
where $\nor_{e,u},\nor_{e,v}$ denote outward pointing unit (normal) vectors.
Hence we obtain:
\begin{proposition} For all $X\in C^1_\finite(\cg)$ we have
\begin{equation}\label{eq:calcdivergence}
\inte{ \nablacalc\cdot X} = \intv{ \wt\nor\cdot X},
\end{equation}
where
$$
(\wt\nor\cdot X)(v) = \cv(v)^{-1} \sum_{e\ni v}a_e\nor_{e,v}\cdot X|_e(v).
$$
\end{proposition}

Equation~\ref{eq:calcdivergence} shows that $\nablacalc$ cannot be
the right notion of a divergence.  Indeed, in analysis the analogue to
$\wt\nor\cdot X$ is integrated over the {\em boundary}, and we do not
wish to consider every vertex of $\cg$ as a boundary point.  Fortunately
the notion of the divergence is essentially forced upon us by previously
fixed concepts.

Let $C^k_\Dirichlet(\cg)$ denote those functions in 
$C^k_\finite(\cg)$ that vanish on the boundary of $\cg$.

\begin{definition} For a vector field, $X$, its {\em divergence functional} 
is the linear functional 
${\cal L}_X\from C^\infty_\Dirichlet(\cg)\to\reals$ given
by,
$$
{\cal L}_X(g) = -\inte{ X \cdot \nabla g} .
$$
\end{definition}

\begin{proposition} For any $X\in C^1(\tcg)$ and 
$g\in C^\infty_\Dirichlet(\cg)$ we have
$$
{\cal L}_X(g) = \inte{(\nablacalc\cdot X)g} - \intv{(\wt\nor\cdot X)g},
$$
i.e. the divergence functional of $X$ is represented by
$(\nablacalc\cdot X)\ce - (\wt\nor\cdot X)\cv$ (viewed as a linear functional
via integration).
\end{proposition}
\proof We substitute $Xg$ for $X$ in equation~\ref{eq:calcdivergence},
and note that $\nablacalc\cdot(Xg) = g\nablacalc\cdot X + X\cdot\nabla g$.

\begin{definition}\label{df:divergence}
For $X\in C^1(\tcg)$ we define its {\em divergence},
$\nabla\cdot X$, to
be the measure
$$
(\nablacalc\cdot X)d\ce - (\wt\nor\cdot X)d\cv.
$$
If $X$ is
edgewise constant, so that $\nablacalc\cdot X=0$, we will also refer to
$$-\wt\nor\cdot X$$ (a function defined only on vertices) as its divergence,
and write $\nabla\cdot X$ for it.
\end{definition}

Definition~\ref{df:divergence} clearly involves some amount of foresight
and/or cheating.  Indeed, since ${\cal L}_X$ is defined only on functions
that vanish on $\partial\cg$, we have no business defining $\nabla\cdot X$
on a boundary vertex.  Pedantically, for $g\in C^1_\finite(\cg)$ we should 
now search
for the missing term in
$$
-\inte{X\cdot\nabla g} = \int_{\cg\setminus\partial\cg}(\nabla\cdot X)g +
\,\mbox{missing term} .
$$
But by equation~\ref{eq:calcdivergence} (substituting $Xg$ for $X$) we have
$$
\mbox{missing term}\, = -\int_{\partial\cg} (\wt\nor\cdot X)g\,d\cv =
\int_{\partial\cg}(\nabla\cdot X)g .
$$
We conclude:
\begin{proposition}\label{pr:divergence}
For any $g\in C^1_\finite(\cg)$ and $X\in C^1_\finite(\tcg)$ we have
$$
\int_\cg (\nabla\cdot X)g + \inte{X\cdot \nabla g\;} = 0.
$$
To make this look more like analysis we can write this as:
$$
\int_{\cg\setminus\partial\cg} (\nabla\cdot X)g + \inte{X\cdot \nabla g} = 
\int_{\partial\cg} (\wt\nor\cdot X)g\,d\cv .
$$
\end{proposition}

\subsection{The Laplacian}

In graph theory we usually define positive semidefinite Laplacians.  So
we define
$$
\Delta f = -\nabla\cdot(\nabla f).
$$

By using 
proposition~\ref{pr:divergence} we obtain
\begin{proposition} For all $f\in C^2_\finite(\cg)$ and 
$g\in C^1_\finite(\cg)$ we have
\begin{equation}\label{eq:laplacian_positive}
\int(\Delta f)g = \inte{\nabla f\cdot\nabla g}.
\end{equation}
If also $g\in C^2_\finite(\cg)$ we have
\begin{equation}\label{eq:laplacian_symmetric}
\int(\Delta f)g = \int(\Delta g)f.
\end{equation}
\end{proposition}

\begin{proposition} For $f\in C^2_\finite(\cg)$ which is edgewise linear
we have $\nablacalc\cdot\nabla f=0$ and so $\Delta f=\wt\nor\cdot\nabla f
\,d\cv$.  Viewing $\Delta f$ as a function on vertices we therefore have:
\begin{equation}\label{eq:laplacian_explicit_old}
(\Delta f)(v) = \cv(v)^{-1}\sum_{e\sim\{u,v\}}a_e\frac{f(v)-f(u)}{\ell_e}.
\end{equation}
\end{proposition}

When restricting to edgewise linear functions, it is common (in graph
theory) to write $\Delta$ as $D-A$, where $D$ is the diagonal matrix or
operator (classically the ``degree'' matrix) whose $v,v$ entry is:
$$
L(v) = \cv(v)^{-1}\sum_{e\sim\{u,v\}}a_e/\ell_e,
$$
where we omit $e$'s that are self-loops from the summation,
and where $A$ is the ``adjacency'' matrix or operator given by
$$
(Af)(v) = \cv(v)^{-1}\sum_{e\sim\{u,v\}}(a_e/\ell_e)f(u),
$$
again omitting self-loops, $e$.

A standard and easy application of Cauchy-Schwartz shows that
$$
|(Af,f)| \le (Df,f)
$$
when $f$ is edgewise linear (allowing for the possibility that one or
both sides is $+\infty$).

We will now view $\Delta$ as an operator, and bound its norm.
To simplify matters, we shall assume for the rest of this subsection that
our graph is locally finite, i.e. each vertex is incident upon only finitely
many edges.

Let
$L^2_\Dirichlet(\cg,\cv)$ be the subspace of edgewise linear
functions which vanish on $\partial\cg$ and which
lie in $L^2(\cg,\cv)$.
We make $\Delta$ operate on $f\in L^2_\Dirichlet(\cg,\cv)$ 
by taking $\Delta f$ to be $0$ on $\cg$, to be defined by
equation~\ref{eq:laplacian_explicit_old} on other vertices, and edgewise
linear; as such, $\Delta f$ may not lie in $L^2(\cg,\cv)$, and in this
case we view $\Delta$ as undefined on $f$.

Consider the norm of $\Delta$ as an operator on $L^2_\Dirichlet(\cg,\cv)$,
$$
\|\Delta\|= \sup_{f\in L^2_\Dirichlet(\cg,\cv)} \|\Delta f\|/\|f\|.
$$
Just as in traditional graph theory (see \cite{mohar_infinite}) we have:

\begin{proposition}
$L_{\sup}\le\|\Delta\|\le 2L_{\sup}$, where
$L_{\sup}$ is the supremum of the $L(v)$ (over $v\notin\cg$).
In particular, $\Delta$ is
bounded iff $L_{\sup}$ is.
\end{proposition}
\proof
For $v\in V$ it is clear that
$$
\|\Delta\chi_v\|/\|\chi_v\| \ge L(v),
$$
where $\chi_v$ is the edgewise linear characteristic function of $v$,
i.e. $1$ on $v$, $0$ on other vertices, and edgewise linear.  Hence the
first inequality.  For the second we have for a function of finite type
$$
(\Delta f,f)\le |(Af,f)|+(Df,f)\le 2(Df,f)\le 2L_{\sup}(f,f).
$$
Since $\Delta$ is symmetric on functions of finite type, and since the
set of such functions is dense in $L^2_\Dirichlet(\cg,cv)$, we conclude
the second inequality.
\proofbox

\subsection{Co-area Formulas}

When an integral involves $|\nabla f|$, this term can be removed by
integrating over the level surfaces of $f$.  The simplest case of this
is the one dimensional case, which intuitively expresses the
first-year calculus formula: $df=f'(x)dx$.  The general case is
referred to as the {\em co-area} formula.  We will use it in our
Federer-Fleming theorems.

Recall that for each $e\in E$ we have an associated weight, $a_e$ (that 
determines $\ce$ restricted to $e$).
\begin{definition} For $x\in\cg\setminus V$ we define the {\em area} of
$x$ to be $a_e$ where $e$ is the edge containing $x$.
For $F\subset\cg\setminus V$ we define its {\em area}, $\area(F)$,
to be the sum of the areas of all non-vertex points in $F$.
\end{definition}
So if $\Omega\subset\cg$, it makes sense to speak of $\area(\partial\Omega)$
provided that $\partial\Omega$ contains no vertices.
Later in this subsection we define $\area(\partial\Omega)$ even when 
$\partial\Omega$ contains vertices.
 
For any $f\in C^0(\cg)$ we set
$$
\Omega(t) = \Omega_f(t) = \{ x\in\cg | f(x)>t\}.
$$
For $f\in \Lipschitz_\finite(\cg)$ we have 
that for almost all $t$, $\partial\Omega(t)$
is a finite set of points (see \cite{morgan}, page 24).

\begin{proposition}[The Co-Area Formula] Let $f\in \Lipschitz_\finite(\cg)$ 
and for any
$t$ let
$\area_t$ be the restriction of $\area$ to
$\partial\Omega_f(t)$.  Then we
have $|\nabla f| \,\de = d\area_t \,dt$, in the sense that
for any $\phi\in C^0(\cg)$ we have 
$$
\inte{\phi\,|\nabla f|}
=\int\biggl(\int\phi\, d\area_t\biggr) \,dt
$$
(where the integral in $t$ is taken to be a Lebesgue integral).
In particular:
$$
\inte{|\nabla f|} = \int\int d\area_t \,dt = \int \area
\Bigl(\partial\Omega(t)\Bigr)\,dt.
$$
\end{proposition}
Since $\partial\Omega(t)$ contains a vertex for only finitely many $t$, it
is irrelevant how we define its area in the proposition.
Versions of this proposition have appeared implicitly in many places, and
this proposition appears explicitly in \cite{bobhou}.

\proof It suffices to prove this theorem when we integrate over any edge.
It this case this is just the standard co-area formula in one dimension
(see \cite{federer} or \cite{morgan} for a proof).
\proofbox
We remark that the one dimensional co-area formula we use is easy to prove
in virtually all our applications.  Indeed, this formula is equivalent
to saying that
if $f\in \Lipschitz[0,1]$ and $a<b$ are reals, then
\begin{equation}\label{eq:explicitcoarea}
\int_{\{x\,:\,a<f(x)<b\}} |\nabla f|\,dx = 
\int_a^b \Bigl(\mbox{\# of $f^{-1}(t)$}\Bigr)\,dt
\end{equation}
In its applications to calculus on graphs, $f$ will typically be
piecewise differentiable and
$f'$ will change signs a finite
number of times; then equation~\ref{eq:explicitcoarea}
follows immediately from $df=f'(x)dx$ (with $t=f$ in the
formula).  However, to prove the above co-area formula for {\em any}
$f\in \Lipschitz[0,1]$ requires a more subtle argument.

We mention that the co-area formula also holds when our functions
have a finite number of discontinuities.  In this case we interpret
$\inte{\phi |\nabla f|}$ for such $f$ to mean that we add $\phi(x)a_e$ times
the jump in $f$ at $x$ for each point of discontinuity of $f$, where $e\in x$,
and if $x$ is a vertex then we add one such contribution for each $e$ meeting
$x$ using the jump along $e$ of $f$ at $x$.  (This
understanding agrees with how we define $\inte{|\nabla f|}$ as $\cm(f)$
in section~\ref{se:ff}.)  However, to make such a formula valid we must
adopt the definition below.
\begin{definition}  For any open $\Omega\subset\cg$, we say {\em $v$ belongs
to the $e$ closure of $\Omega$}, written $v\in\widebar{\Omega|_e}$,
if the closure of $\Omega$ intersected with
the edge interior of $e$ contains $v$.  We define
$$
\area(\partial\Omega) = \area(\partial\Omega\setminus V) +
\sum_{v\in \partial\Omega\cap V}
\left( \sum_{e \;{\rm with}\; v\in\widebar{\Omega|_e}} a_e \right)
$$
\end{definition}
Clearly $\area(\partial\Omega)$ is just $\inte{|\nabla \chi_\Omega|}$ in
the above sense; as we shall see, it is also $\cm(\chi_\Omega)$ in the sense
of section~\ref{se:ff}.

\begin{proposition} The above co-area formula holds for any finitely
supported locally Lipschitz function with at most a finite number of
discontinuities, given the above definition of area and understanding
of $\inte{\phi |\nabla f|}$.
\end{proposition}
\proof  For each jump we glue in an interval of arbitrary length, and
let us redefine $f$ on each interval to
vary linearly between the values of its two
limits at the jump.  Now apply the old co-area formula to the new graph
and new $f$.
\proofbox

We finish by remarking that it is easy to see that
$$
\area(\partial\Omega) = \lim_{h\to 0} \frac{\ce(\Omega)-\ce(\Omega_h)}{h},
$$
where $\Omega_h$ denotes the set of points in $\Omega$ whose distance to
$\partial\Omega$ is more than $h$.  If we let $\Omega^h$ be the set of points
in $\cg$ whose distance to $\Omega$ is less than $h$ (including all points
in $\Omega$), then \cite{bobhou} give a general co-area formula where the
``area'' would be given by
$$
\lim_{h\to 0} \frac{\ce(\Omega^h)-\ce(\Omega)}{h}.
$$
This definition
disagrees with ours only when $\partial\Omega$ contains vertices, and
this disagreement is important to a co-area formula only when the function
has a discontinuity at one or more vertices.

\subsection{Countability}

We close with some remarks about the countability of our graphs.
Nowhere do we assume that our graphs have countable vertex sets, edges sets,
or in particular vertex
degrees.  However, for the sake of intuition,
in most applications it is fairly safe to assume
that the vertex and edge sets and vertex degrees are countable.
For example, if $\|f\|_q,\|\nabla f\|_p$ are finite for some $p,q$, then 
clearly $f$ vanishes
at any vertex of uncountable degree; such vertices may exist, but they tend
to be forced ``boundary points''  (so neighbours of such points can't
``interact'' through that point).  Also, notice that if each vertex has
countable degree, then each connected component of $\cg$ has only countably
many vertices and edges.

Notice that there are many important graphs with some countably
infinite 
vertex degrees, such as those in \cite{adler}.  However, for those graphs
the lengths of the edges incident upon a vertex
tend to infinity essentially as a geometric series,
and so the Laplacians, for example, are still finite.

\section{Preliminaries for Gradient Inequalities}

By a {\em gradient inequality} we mean a lower bound on $\|\nabla f\|_p$
in terms of norms on $f$ and constants that may depend on $\cg$.
In this section we give some preliminary concepts needed to state and
prove the gradient inequalities appearing in the rest of this paper.

\subsection{Edgewise linearity}

\begin{proposition}\label{pr:linear}
Consider those $f\in C^1_\finite(\cg)$ taking on prescribed
values at all vertices.  For $p>1$, $\|\nabla f\|_p$ is minimized exactly when
$f$ is edgewise linear, and for $p=1$ when $f$ is monotone along each
edge.
\end{proposition}
This proposition tells us that in proving a gradient inequality we can
restrict ourselves to edgewise linear functions.

\proof Clearly for an edge $e=\{u,v\}$ we have
\begin{equation}\label{eq:variation}
\int_e |\nabla f|\,\de \ge |f(u)-f(v)| a_e,
\end{equation}
with equality iff $f$ is monotone.  This proves the statement for $p=1$
(summing over $e$).
For $p>1$ we apply Jensen's or H\"older's
inequality to equation~\ref{eq:variation} to
conclude that the integral
of $|\nabla f|^p$ over $e$ is minimized precisely when $|\nabla f|$ is
constant, i.e. $f$ is edgewise linear.
\proofbox

\subsection{Closed Graphs: Finite Graphs Without Boundary}

When constant functions lie in $C^1_\Dirichlet(\cg)$, there are no
interesting gradient inequalities that hold over all of $C^1_\Dirichlet(\cg)$.
This is the case for finite graphs without boundary (and compact manifolds
without boundary).
\begin{definition} A {\em closed} graph is a finite graph without boundary.
\end{definition}
We form interesting inequalities by working ``modulo constant functions,''
in a sense to made precise below.

Our remarks are very general, and there is no loss in generality in working
with an arbitrary measure, $\mu$, on a space $M$ whose total measure is
finite and non-zero.

For any $r>0$ and $X\in\reals^n$
let $\ex{X}{r}$ denote $|X|^{r-1}X$ (interpreted as $0$ if $X=0$).

\begin{proposition} \label{pr:uniquea}
Let $1<p\le\infty$.  For $f\in L^p(M,\mu)$ there is a unique
$a\in\reals$ such that $\|f-a\|_p<\|f-b\|_p$ for all $b\in\reals$ with
$b\ne a$.  Furthermore, $a$
is the unique solution to
$$
\int \ex{f-a}{p-1}\,d\mu = 0
$$
for $p<\infty$, and for $p=\infty$ we have that $a$ is the average of the
essential supremum and essential infimum of $f$.
\end{proposition}

\proof For $p=\infty$ this is clear, so assume $p<\infty$.
It is easily checked that the
function $g(t)=\|f-t\|_p^p$, for $t\in\reals$, is differentiable, and that
$$
g'(t) = p\int \ex{f-t}{p-1}\,d\mu .
$$
It follows that $g'(t)$
is continuous, strictly increasing, and tends
to $\pm\infty$ as $t\to\pm\infty$.
\proofbox

\begin{definition} $f$ is said to be $p$-balanced if $\|f\|_p\le
\|f-a\|_p$ for any $a\in\reals$.
\end{definition}

The above proposition has a $p=1$ version, although $a$ is not unique.
This version is linked to the important notion of being {\em split}.
\begin{definition} A measurable $f$
is said to be {\em split} if 
$\mu(\{x|f(x)>0\})$ and $\mu(\{x|f(x)<0\})$ are both $\le \mu(M)/2$.
\end{definition}

\begin{proposition} For any $f\in L^1(M)$ let $I\subset\reals$ be the set 
on which
$g(t)=\|f-t\|_1$ achieves its minimum,
and let $J\subset\reals$ be the set
of $t$ such that $f-t$ is split.  Then $I$ and $J$ are nonempty compact
intervals,
and $I\subset J$.
\end{proposition}

\proof $g(t)\ge \|t\|_1-\|f\|_1=|t|\mu(M)-\|f\|_1$.
It follows that $g(t)\to\infty$ as $t\to\pm\infty$.
Since $g(t)$ is continuous (indeed, $|g(t)-g(s)|\le |t-s|\mu(M)$ by the
triangle ineqality), and
convex (using the triangle inequality), it
follows that $I$ is a nonempty compact interval.  Clearly $J$ is a nonempty
compact interval.  Next, if $t\notin J$, then we claim $t\notin I$;
indeed, assume that $\mu(\{x|f(x)>t\})>\mu(M)/2$.  Then there is a
$t_0$ for which $\mu(\{x|f(x)>t\})>\mu(\{x|f(x)>t_0\})\ge \mu(M)/2$,
and it is easy to see that $g(t_0)<g(t)$ so that $t\notin I$.
Similarly if $\mu(\{x|f(x)<t\})>\mu(M)/2$ we similarly conclude $t\notin I$.
\proofbox

\begin{proposition} For any $1\le p\le\infty$ and $f\in L^p=L^p(M,\mu)$
we have
\begin{equation}\label{eq:functional}
\inf_{a\in\reals} \|f-a\|_p = \sup_{\|g\|_{p'}=1,\;\int g=0} \int fg\,d\mu.
\end{equation}
\end{proposition}
\proof Let $a$ minimize $\|f-t\|_p$ as a function of $t$.
For any $g$ as above we have
\begin{equation}\label{eq:longineq}
\int fg\,d\mu = \int (f-a)g\,d\mu
\le \|f-a\|_p \| g \|_{p'} = \|f-a\|_p.
\end{equation}
Thus equation~\ref{eq:functional} holds with $=$ replaced by $\ge$, and
it remains to prove $\le$.

First consider the case $1<p<\infty$.
Equality will hold in equation~\ref{eq:longineq} provided that 
(1) $|g|^{p'}$ is a constant times
$|f-a|^p$, and (2) $g$ and $f-a$ have the same sign.  But (1) just means
that $|g|$ is proportional to $|f-a|^{p/p'}=|f-a|^{p-1}$; so taking
$g$ proportional to $\ex{f-a}{p-1}$ in such a way that $\|g\|_{p'}=1$,
proposition~\ref{pr:uniquea} shows that $\int g=0$ and thus we conclude
the proposition.

For $p=1,\infty$ we need special arguments to construct $g$ as above.
Consider $p=1$.
Take $A,B$ respectively to be the sets where $f-a$ is
positive and negative, and set
$$
g(x) = \left\{ \begin{array}{ll} 1 & \mbox{if $x\in A$,} \\
				-1 & \mbox{if $x\in B$,} \\
				c & \mbox{otherwise,}  \end{array}\right.
$$
where $c$ is any number between $-1$ and $1$ if $\mu(A)=\mu(B)=\mu(M)/2$,
and otherwise
$$
c = \frac{\mu(A)-\mu(B)}{\mu(M)-\mu(A)-\mu(B)}
$$
(since $f-a$ is split, we have $\mu(M)-\mu(A)-\mu(B)>0$ unless
$\mu(A)=\mu(B)=\mu(M)/2$).
If $\mu(A)=\mu(B)=\mu(M)/2$, then clearly $\int g\,d\mu=0$ and $\|g\|_\infty
=1$, and $\int fg\,d\mu=\|f-a\|_1$.  If not, then $\mu(M)-2\mu(B)>0$ since
$f-a$ is split, and so
$$
c+1 =  \frac{\mu(M)-2\mu(B)}{\mu(M)-\mu(A)-\mu(B)} >0.
$$
Similarly $c-1<0$, and we conclude $-1<c<1$.  Thus $\|g\|_\infty=1$, and it
is easy to check that $\int g\,d\mu=0$ and $\int fg\,d\mu=\|f-a\|_1$.

Next consider $p=\infty$.  There is no analogue of $g$ in this case, but
we claim there is a sequence, $g_\epsilon$, defined for small $\epsilon>0$,
such that $\int fg_\epsilon\,d\mu\to\|f-a\|_\infty$.
Namely, for any $\epsilon>0$ let $A_\epsilon$ be the set where $f+\epsilon>$
than the essential supremum of $f$, and $B_\epsilon$ the set where 
$f-\epsilon$ is small than the essential infimum.  Set
$$
g_\epsilon = \frac{1}{2\mu(A_\epsilon)}\chi_{A_\epsilon} -
\frac{1}{2\mu(B_\epsilon)}\chi_{B_\epsilon}.
$$
It is easy to see that $\|g_\epsilon\|_1=1$, that $\int g_\epsilon\,d\mu=0$,
and that $\int fg_\epsilon\,d\mu$ is within $\epsilon$ of $\|f-a\|_\infty$.
\proofbox
\section{Federer-Fleming Theorems}\label{se:ff}

In this section we prove some Federer-Fleming type theorems.  
Roughly speaking,
these theorems say that certain functionals attain their 
minimum on characteristic functions.  
There are many approaches to proving such theorems; our approach most
closely follows that of Rothaus in \cite{Rotha}.
In the first two subsections we state the theorems.
Then we
give an overview of how these theorems are proved, based on a simple 
inequality.  The later subsections give the details.

\subsection{Statement of the Federer-Fleming Theorem}

The classical Federer-Fleming Theorem looks as follows in graph theory.
By an {\em admissible} $\Omega\subset\cg$ we mean an open $\Omega$ with
$\partial\Omega$ finite and $\Omega$ disjoint from $\partial\cg$.
For $1\le \nu\le \infty$ we set
$$
i_\nu(\Omega) = \area(\partial \Omega)/\cv(\Omega)^{1/\nu'}, \qquad
I_\nu(\cg) = \inf_{\Omega\;{\rm admissible}} 
i_\nu(\Omega)
$$
(if $\ce(\Omega)=0$ we take $i_\nu(\Omega)=+\infty)$.
Next we set
$$
s_\nu(f) = \| \nabla f\|_1 / \| f \|_{\nu'}, \qquad
S_\nu(\cg) = \inf_{f\subset C^1_\Dirichlet(\cg)} s_\nu(f),
$$
where we understand the norms on $\nabla f$ and $f$ to be with respect
to, respectively, $\ce$ and $\cv$.

We shall prove in this section the graph theoretic analogue of the
Federer-Fleming theorem:
\begin{theorem} For any $1\le\nu\le\infty$ we have
$I_\nu(\cg)=S_\nu(\cg)$.
\end{theorem}
This was essentially proven\footnote{Although $I=S$ was not stated in either
article, the proofs of weaker statements given there are easily modified to
give $I=S$; for example, in \cite{chavel-feldman} lemma 4 involves a 
constant depending on $\nu$; but this constant can be taken to be $1$, as
a standard inequality
(equation (6.11),
page 269, of \cite{chavel-intro}) shows; 
this gives $I=S$ (or $2I=S$ with their
conventions).}
in \cite{varopoulos,chavel-feldman}.

The inequality $S\le I$ follows easily by (any reasonable way of) 
approximating any characteristic
function of finite type by $C^1_\Dirichlet$ functions.  The reverse
inequality is discussed in the following few subsections.

The Federer-Fleming theorem above can be viewed as a gradient inequality:
\begin{corollary}\label{cr:split} For any $f\in C^1_\Dirichlet(\cg)$ we have
$$
\| \nabla f\|_1 \ge I_\nu(M) \| f \|_{\nu'}.
$$
\end{corollary}

\subsection{Federer-Fleming for Closed Graphs}

The Federer-Fleming
theorem above can be interesting when $\cg$ is not finite or has a boundary.
However, for finite graphs without boundary, this theorem just says that
$0=0$, by considering $f=1$ and $\Omega=\cg$.  

The traditional way to remedy the fact that
$I_\nu=S_\nu=0$ in the closed case is to set
$$
\wt i_\nu(\Omega) = |\partial \Omega| \min(|\Omega|,
|\widebar \Omega|)^{(1/\nu)-1},
\qquad  \wt I_\nu(\cg) = \inf_{\Omega\;{\rm admissible}} \wt i_\nu(\Omega),
$$
and
$$
\wt S_\nu(\cg) =  \inf_{f\subset C^1_\Dirichlet(\cg)} \sup_{a\in\reals} 
\wt s_\nu(f-a)
=\inf_{f\subset C^1_\Dirichlet(\cg)} \frac{\|\nabla f\|_1}{\min_{a\in\reals}
\|f-a\|_{\nu'}}
$$
One can prove a Federer-Fleming-type inequality, namely:
\begin{theorem}[Closed Federer-Fleming, traditional] If $1\le\nu\le\infty$,
we have
$\wt I_\nu(\cg)\le \wt S_\nu(\cg) \le 2^{1/\nu} \wt I_\nu(\cg)$.
\end{theorem}

While the above definitions of $\wt I_\nu$ and $\wt S_\nu$ are perhaps the most
natural closed case versions of $I_\nu$ and $S_\nu$, 
more precise inequalities are obtained with the following
definitions.

Let
$$
\wt i'_\nu(\Omega) = |\partial \Omega| 
(|\Omega|^{1-\nu}+|\complement\Omega|^{1-\nu})^{1/\nu}
\qquad  \wt I'_\nu(\cg) = \inf_{\Omega\;{\rm admissible}} \wt i'_\nu(\Omega).
$$
Clearly $\wt i_\nu(\Omega)\le \wt i'_\nu(\Omega)\le 
2^{1/\nu}\wt i_\nu(\Omega)$, so
$\wt I_\nu\le \wt I'_\nu\le 2^{1/\nu}\wt I_\nu$.
Let
$$
\wt S_{',\nu}=\inf_{f\in C^1_\Dirichlet(\cg),\;f \rm split} s_\nu(f).
$$
Our more precise version of compact Federer-Fleming is:

\begin{theorem}\label{th:ourff} For any $1\le \nu\le\infty$ we have
$\wt I'_\nu=\wt S_\nu$ and $\wt I=\wt S_{',\nu}$ 
(and clearly  $\wt I_\nu\le \wt I'_\nu\le 2^{1/\nu}\wt I_\nu$).
\end{theorem}
The following important corollary follows:
\begin{corollary} For $f$ split we have
\begin{equation}\label{eq:estimate}
\|\nabla f \|_1 \ge \wt I_\nu(\cg) \| f \|_{\nu'}.
\end{equation}
\end{corollary}
The above corollary is much easier to work in applications to Sobolev
inequalities
than the weaker claim that for any $f$ we have
$$
\|\nabla f \|_1 \ge \wt I_\nu(\cg) \| f\|_{\nu'}
$$
for $f$ $\nu'$-balanced.
The reason for this is that if $f$ is split, then
so is $\ex{f}{\alpha}$; this is not true for $f$ balanced.

We finish by remarking that if we try to interpret what
$\wt I,\wt I',\wt S,\wt S_{'}$ would mean on a $\cg$ with $\cv(\cg)=\infty$
or with a boundary, then we claim we would recover
$I,S$.  So the closed versions
of the Federer-Fleming theorems proven here are reasonable analogues of the
traditional Federer-Fleming theorems.

\subsection{A Simple Inequality}

If $M,L\in L^1[-T,T]$ are non-negative functions, then we have
\begin{equation}\label{eq:verysimple}
\frac{\int_{-T}^T M(t)\,dt}{\int_{-T}^T L(t)\,dt}
\ge \inf_{t\in[-T,T]} \frac{M(t)}{L(t)}
\end{equation}
(with the convention that $a/0$ is $+\infty$ for any $a\ge 0$).
In other words, the quotient of two superpositions (or averages) is at least
as big as the minimum of the individual quotients.  In the rest of this
subsection we discuss mild variants of this very simple inequality.

If $\cf$ is a normed vector space, for a family $\{ f_t\}_{t=-T}^{t=T}$ of
elements of $\cf$ it makes sense to write ``superpositions'' or integrals
$$
f = \int_{-T}^T f_t\,dt
$$
as the limit (in the norm) of Riemann sums, presuming the limit exists.
We say that a functional $\ch\from\cf\to\reals$ is subadditive (respectively
additive) on a superposition as above if
$$
\ch(f) \le \int_{-T}^T \ch(f_t)\,dt
$$
(respectively equality holds).  So if $\cq=\cm/\cl$ is a quotient of two
non-negative functionals, with $\cm$ additive on the above superposition
and $\cl$ subadditive, we have
$$
\cq(f) \ge \min_{t\in[-T,T]} \cq(f_t),
$$
using equation~\ref{eq:verysimple}.

Now any $f\in C^1_\Dirichlet(\cg)$ can be written as a superposition of 
$\pm\chi_\Omega$, characteristic and negative characteristic functions,
almost all $\Omega$ having finite boundary; here
the superposition understood in some appropriate normed vector space, $\cf$,
containing $C^1_{\Dirichlet}(\cg)$ and all the characteristic functions, such as
$L^p(\cg)$.  From the above discussion we see:
\begin{proposition}\label{pr:Q=M/L}
Let $\cq=\cm/\cl$ be a quotient of two
non-negative functionals on a space, $\cf$, as above.  Assume that we can
write
any $f\in C^1_\Dirichlet(\cg)$ as a superposition of plus/minus 
characteristic functions
on which $\cm$ and $\cl$ respectively additive and
subadditive.  Then
$$
\min_{f\in C^1_\Dirichlet(\cg)}  \cq(f) \ge  
\min_{\Omega\;\rm admissible}
\cq(\pm\chi_\Omega).
$$
\end{proposition}
In fact, we will see that the above inequality holds with equality for the
$\cm$'s and $\cl$'s of interest to us.

\subsection{Our $\cm$}

In this paper we will be concerned with only one functional, $\cm$.  Namely,
$$
\cm(f) = \sup_{\mbox{$X\in C^1_\finite(\tcg)$, $|X|\le 1$}} 
-\int (\nabla\cdot X)f.
$$
This functional is defined on $L^p(\cg)$ for any $1\le p\le\infty$ and on
$C^k(\cg)$ for any $k\ge 0$; however, it can take on the value $+\infty$.
\begin{proposition} $\cm$ satisfies the following properties:
\begin{enumerate} 
\item for $f\in C^1_\finite(\cg)$ we have $\cm(f)=\inte{|\nabla f|}$,
\item for open $\Omega\subset\cg$ with finite boundary, we have
$\cm(\pm \chi_\Omega) = \area(\partial\Omega)$, and
\item for $f\in C^1_\finite(\cg)$ with $|f|\le T$, we have that 
$\cm$ is additive
with respect to the superposition
$$
f = \int_0^T \Omega_f(t)\,dt + \int_0^T -\Omega_{-f}(t)\,dt
$$
(with $\Omega_f(t)=\{ x\in\cg|f(x)>t\}$ as before).
\end{enumerate}
\end{proposition}

\proof By the divergence theorem we have that if $f\in C^1_\finite(\cg)$ then
$$
\cm(f) = \sup_{X\; \rm as\; above} \inte{X\cdot\nabla f},
$$
and it follows that $\cm(f)\le\inte{|\nabla f|}$.  The reverse inequality is
done by approximating $\nabla f/|\nabla f|$.  Namely, for any $\epsilon>0$,
the set where $|\nabla f|>\epsilon$ is open, and it therefore contains a finite
boundary open subset, $\Omega_\epsilon$ covering all but an $\ce$-measure 
$\epsilon$ set of its points.  Then
$$
\int_{\cg\setminus\Omega_\epsilon}|\nabla f|\,d\ce
\le \epsilon\ce\Bigr(\supp(f)\Bigr)
+ \epsilon\sup|\nabla f|
$$
which tends to zero as $\epsilon\to 0$.  Now take $X_\epsilon
=\nabla f/|\nabla f|$
on $\Omega_\epsilon\setminus V$, and otherwise anything less than one in norm
in a smooth way.  Then $X_\epsilon\in C^\infty_\finite(\tcg)$ and
$$
\inte{X\cdot\nabla f} \ge \int_{\Omega_\epsilon} |\nabla f|\,d\ce
- \int_{\cg\setminus\Omega_\epsilon} |\nabla f|\,d\ce
= \inte{|\nabla f|} - 2 \int_{\cg\setminus\Omega_\epsilon} |\nabla f|\,d\ce
$$
and the last expression tends to $\inte{|\nabla f|}$.  Thus the reverse
inequality is proven.

The second statement is proven by the divergence theorem applied to
$\Omega$, namely
$$
\cm(\chi_\Omega) = \sup_{X\; \rm as\; above} -\int_{\partial\Omega}
\wt\nor\cdot X.
$$
This makes $\cm(\chi_\Omega)\le\area(\partial\Omega)$ clear.  For the reverse
inequality we take 
$X$ to have $|X|=1$ and
to be of the right direction in a neighbourhood of its boundary, and anything
elsewhere provided that $|X|\le 1$ everywhere and $X\in C^1_\finite(\tcg)$.

The additivity follows immediately from the co-area formula applied to
$f^+=\max(f,0)$ and $f^-=\max(-f,0)$.

\subsection{Our $\cl$'s: Sup-linear functionals}

We next describe the $\cl$'s that we consider.
\begin{definition} Let $\cf$ be a normed linear space, and let
$H$ be a collection of bounded, linear functionals on $\cf$.  We define
the {\em sup-linear functional associated with $H$} to be
$$
\ch(f) = \sup_{\ell\in H} \ell(f).
$$
\end{definition}
We now record some simple remarks:
\begin{proposition}\label{pr:L}
The following hold:
\begin{enumerate}
\item If $0\in H$, then $\ch$ is non-negative.  
\item If $H$ is
symmetric, i.e. $\ell\in H$ implies $-\ell\in H$, then $\ch(f)=\ch(-f)$.
\item $\ch(f_1+f_2)\le \ch(f_1)+\ch(f_2)$ for any $f_i\in\cf$.
\item $\ch(\alpha f) = \alpha\ch(f)$ if $\alpha$ is any positive real.
\item If $f_n\to f$ in $\cf$, then $\ch(f)\le \liminf\ch(f_n)$ (since the
supremum of lower semicontinuous functions is lower semicontinuous).  
In particular, using 3 and 4,
$\ch$ is subadditive with respect to {\em any} superposition in $\cf$.
\item If $H$ is uniformly bounded, then $\ch(f)$ is continuous.
\end{enumerate}
\end{proposition}

Now we give some examples of sup-linear functionals:
\begin{enumerate}
\item $\cm(f)$ as defined in the previous subsection.
\item $\cl(f)$ being the $L^p(\cg,\cv)$ norm, since then
$$
\cl(f) = \sup_{\|g\|_{p'}=1}\intv{gf},
$$
\item
$$
\cl(f) = \inf_{a\in\reals} \|f-a\|_{p,\cv} = 
\sup_{\|g\|_{p'}=1,\;\intv{g}=0}\intv{gf},
$$
\end{enumerate}
The functionals mentioned in items 2 and 3 are continuous with respect to the
$L^p(\cg,\cv)$ norm, in view of proposition~\ref{pr:L}, item 6.  More 
interesting examples are given in \cite{Rotha}.

\subsection{Approximating $\chi_\Omega$}

In proposition~\ref{pr:Q=M/L} it is usually easy to see that equality
holds.  The reason for this is that if $\Omega$ is admissible, then
one can approximate $\chi_\Omega$ by a sequence of $C^1_\Dirichlet(\cg)$
functions.  We make this precise here.

So fix an admissible $\Omega\subset\cg$.  For any $\epsilon>0$ sufficiently
small, let
$f_\epsilon$ be a function such that
\begin{enumerate}
\item $f_\epsilon$ vanishes on $\complement\Omega$,
\item $f_\epsilon=1$ on the set of points, $A_\epsilon$, 
whose distance to $\complement\Omega$
is at least $\epsilon$, and
\item along any edge segment of length $\epsilon$ joining $\complement\Omega$
and $A_\epsilon$, $f_\epsilon$ increases monotonically from $0$ to $1$ with
$f_\epsilon\in C^1(\cg)$.
\end{enumerate}
Then we have:
\begin{enumerate}
\item $\cm(f_\epsilon)=|\partial\Omega|$ for $\epsilon$ sufficiently small.
\item $f_\epsilon\to\chi_\Omega$ as $\epsilon\to 0$ in $L^p(\cg,\cv)$ for
any $1\le p<\infty$.
\item $\cl(f_\epsilon)\to\cl(\chi_\Omega)$ for any of the $\cl$'s mentioned
in the previous sections.  (By the previous remark this holds for the
$\cl$'s with $p<\infty$, and it is easy to verify the $p=\infty$ case
directly.)
\end{enumerate}
Notice that the above construction works for any open $\Omega$ of finite
type, even if $\partial\Omega$ is infinite.
In this case items 2 and 3 are still true,
and while item 1 is not true, we still have
$\cm(f_\epsilon)\to|\partial\Omega|$.

We easily conclude:
\begin{proposition} For $\cm$ as described above, and any $\cl$ mentioned
in the previous section we have
$$
\min_{f\in C^1_\Dirichlet(\cg)}  \cq(f) =  
\min_{\Omega\;\rm admissible}
\cq(\pm\chi_\Omega).
$$
\end{proposition}

\subsection{Proofs of the Federer-Fleming statements}

The Federer-Fleming statements follow easily from the proceding
discussions.  For the traditional Federer-Fleming theorem, we have
$S_\nu= I_\nu$ by the previous discussion, taking
$\cl(f)$ to be the $L^{\nu'}(\cg,\cv)$ norm, seeing as then
$$
\cl(\chi_\Omega)= |\Omega|^{1/\nu'}.
$$

For closed graphs, $\wt I_\nu\le \wt S_{',\nu}$ 
follows since for $f$ split
we have that the superposition of $f$ as characteristic functions involves
only $\chi_\Omega$ with $|\Omega|\le\cv(\cg)/2$.  The reverse inequality
follows by the approximation argument in the previous section, since
$|\Omega|\le\cv(\cg)/2$ implies that the $f_\epsilon$ in the previous
section are split.

To see
$\wt I'_\nu= \wt S_\nu$, we consider
$\cl(f)=\min_{a\in\reals}\|f-a\|_{\nu'}$.  For
any open $\Omega$ of finite type we have the minimum of 
$\|\chi_\Omega-a\|_{\nu'}$ is attained when
$$
\intv{\ex{\chi_\Omega-a}{\nu'-1}} = 0,
$$
i.e. when
$$
(1-a)^{\nu'-1}|\Omega|=a^{\nu'-1}|\complement\Omega|,
$$
i.e. for
$$
a = \frac{|\complement\Omega|^{\nu-1}}{|\Omega|^{\nu-1}+
|\complement\Omega|^{\nu-1}}.
$$
It follows that 
$$
\cl(\chi_\Omega)=(|\Omega|^{1-\nu}+|\complement\Omega|^{1-\nu})^{1/\nu},
$$
and thus
$(\cm/\cl)(\chi_\Omega)=i'_\nu(\Omega)$.  Thus 
$\wt I'_\nu= \wt S_\nu$.

\subsection{The Generalization of Rothaus}

In the paper \cite{Rotha} of Rothaus, it is noted that $\cl$ may be
generalized to
$$
\cl(f) = \cl_1(f^+) + \cl_2(f^-),
$$
where $f^+=\max(f,0)$, $f^-=\max(-f,0)$, and $\cl_i$ for $i=1,2$ are 
functionals as before.  It is also shown there that if we further
restrict our
functions from $C^1_\Dirichlet(\cg)$ to those which in addition satisfy
$$
\intv{f^+P}=\intv{f^-Q},
$$
where $P,Q$ are fixed positive function on $\cg$, then we get a
similar Federer-Fleming theorem, where $\pm\chi_\Omega$ is replaced by
$$
Q(B)\chi_A-P(A)\chi_B,
$$
where $A,B$ range over all admissible subsets of $\cg$ with disjoint
closures, and $Q(B),P(A)$ are the integrals $d\cv$ of $Q,P$ respectively
over $B,A$.

Rothaus wrote his paper for manifolds, but all the proofs immediately
carry over to the graph theory case using our framework.

\subsection{Remarks on the Isoperimetric Constants}

In graph theory one often defines the isoperimetric constants (i.e. our
$I_\nu,\wt I_\nu,\wt I'_\nu$) in terms
of subsets of vertices and the edges leaving them.  Since we define our
constants as infimums of certain quantities over all {\em admissible} sets,
one might wonder if our constants agree with classical ones in graph
theory.  In fact, it is easy to see that they do.

\begin{proposition} In defining any one of $I_\nu,\wt I_\nu,\wt I'_\nu$,
we may restrict the admissible sets, $\Omega$ by further requiring that 
(1) $\Omega$ is connected, and (2) if $x,y\in\Omega$ lie on an edge interior
or its closure, then the entire edge segment from $x$ to $y$ lies in $\Omega$.
\end{proposition}
Any of $i_\nu,\wt i_\nu,\wt i'_\nu$ of an $\Omega$ as above will be
determined by which vertices lie in $\Omega$, and agree with their classical
graph theory analogues.
\proof Part (1) is an observation of Yau (in analysis).  Indeed, if 
$A,B$ are admissible and disjoint, then clearly
$\chi_A+\chi_B$ is a superposition for which our $\cm$ is additive.
It follows that any of  $I_\nu,\wt I_\nu,\wt I'_\nu$ are at least as small on
$\chi_A$ or $\chi_B$ as they are on the sum.  Part (2) follows from the
fact that if we add the edge segment from $x$ to $y$ to $\Omega$, then
$\cm$ can't increase and $\cl$ doesn't change.
\proofbox

\section{Dodziuk's and Alon's Versions of Cheeger's Inequality}

We briefly described how Dodziuk's and Alon's versions of Cheeger's
inequality can be derived in our framework, their similarities, and
their differences.

Let
$$
\lambda = \inf_{f\in C^1_\Dirichlet(\cg)} {\cal R}(f),\quad\mbox{where}\quad
{\cal R}(f)=\frac{\inte{|\nabla f|^2}}{\intv{f^2}}.
$$
This $\lambda$ can be understood as the first Laplacian eigenvalue, at least
when the infimum is achieved by some $f$, and is of fundamental importance
in spectral theory.  We wish to lower bound $\lambda$.

Notice that when $\cg$ is closed we have $\lambda=0={\cal R}(1)$.  At the
end of this section we will
modify our notion of $\lambda$ in this case and state the analogous theorems.

\subsection{The Basic Technique}

To bound $\lambda$ from below
we notice that for $X\in C^1_\finite(\cg)$ and 
$f\in C^1_\Dirichlet(\cg)$ we have, using Cauchy-Schwartz,
$$
\inte{X\cdot\nabla(f^2)} = \inte{2fX\cdot\nabla f}\le 2\|fX\|_{2,\ce}
\|\nabla f\|_{2,\ce}.
$$
Dividing by $\intv{f^2}$ yields
$$
\cq_1(f)\le 2\cq_2(f)\sqrt{{\cal R}(f)},
$$
where
$$
\cq_1(f) = \frac{\inte{X\cdot\nabla(f^2)}}{\intv{f^2}},\quad\mbox{and}\quad
\cq_2(f) = \frac{\|fX\|_{2,\ce}}{\|f\|_{2,\cv}}.
$$
Hence we have:
\begin{proposition}\label{pr:basic_technique}
For any $f\in C^1_\Dirichlet(\cg)$ and
$X\in C^1_\finite(\cg)$ and let $\cq_i(f)=\cq_i(f,X)$ be as
above.  Then
$$
\lambda \ge \inf_f\sup_X \cq_1^2(f,X)/\Bigl(4\cq_2^2(f,X)\Bigr).
$$
If we understand that for each $f$ we have specified an $X$, then
$\cq_i(f,X)=\cq_i(f)$ and we have, in particular,
$$
\lambda \ge \inf_f \cq_1^2(f)/\Bigl(4\cq_2^2(f)\Bigr).
$$
\end{proposition}

Of course, in the closed case we have $\lambda=0$ and $\cq_1(f)=0$ for
$f$ constant; what we seek
in this case
is a lower bound on the {\em second} Laplacian eigenvalue.  We discuss
this case in a later subsection.

\subsection{Dodziuk's Lower Bound}

Dodziuk takes $X=\nabla f/|\nabla f|$ when $\nabla f\ne 0$, and $X=0$
otherwise.  Then
$$
\cq_1(f)=\frac{\inte{|\nabla(f^2)|}}{\intv{f^2}} \ge I_\infty(\cg).
$$
Furthermore
\begin{equation}\label{eq:cq2}
\cq_2(f)=\|f\|_{2,\ce}/\|f\|_{2,\cv},
\end{equation}
and according to equation~\ref{eq:rhoconcaveineq} this is bounded
above by $\rho_{\sup}^{1/2}$.  We conclude:
\begin{theorem}[Dodziuk] $\lambda\ge I_\infty^2/(4\rho_{\sup})$.
\end{theorem}

\subsection{Alon's Lower Bound}

Rather than use $I_\infty(\cg)$ to lower bound $\lambda$, Alon uses what
he calls the {\em magnification}, a type of ``vertex expansion'' that
arises in many computer science settings including
sorting and communication networks.

\begin{definition} $\cg$ is a $c$-magnifier if for subset of vertices,
$A\subset\cg\setminus\partial\cg$ we have $|\Gamma(A)|\ge(1+c)|A|$, where
$\Gamma(A)$ denotes the neighbours of $A$, i.e. those vertices (possibly
belonging to $A$) with an edge joining them to $A$.
\end{definition}

Alon's choice of $X$ (in the sense of proposition~\ref{pr:basic_technique})
is more involved.  For the moment assume $\cv(v)=1$ for all $v\in V$ and
$a_e=1$ for all $e\in E$.

\begin{theorem} For any set of vertices, $A\subset\cg\setminus\partial\cg$ 
there exists an $X\in 
C^1_\finite(\cg)$ such that
\begin{enumerate}
\item $|X|\le 1$ everywhere,
\item $-\nabla\cdot X\ge c$ on $A$,
\item $-\nabla\cdot X\le 0$ on $\complement A$, and
\item we have
$\rho_{\sup}(\cg^{|X|^2})\le (2+c^2)\sup_e\ell_e/2$ (and we 
can replace $2+c^2$ with $2+\lfloor c\rfloor + (c-\lfloor c\rfloor)^2$, which
is interesting for $c\ge 1$).
\end{enumerate}
\end{theorem}
\proof 
Form a network with vertices $\{s,t\}\cup B_1\cup B_2$
as follows.  Let $B_1,B_2$ be copies of $A,V$ respectively.
Form an edge of capacity $1+c$
from the source, $s$, to each 
$B_1$.  Form an edge of capacity $1$ from vertex of $B_2$ to the sink,
$t$.  For each $(b_1,b_2)\in B_1\times B_2$ form an edge of capacity $1$
if either $b_2=b_1$ or $\{b_1,b_2\}$ is an edge in $\cg$.  It is not hard
to see (see \cite{alon}) that restricting any
max flow of this network to $B_1\times B_2$ we get an edgewise constant
vector field, $X$, on $\cg$ that satisfies
\begin{enumerate}
\item $|X|\le 1$ everywhere,
\item for any $v$ we have $\sum_{e=\{u,v\}\in E} X^+(v;e) \le 1$, where
$X^+(v;e)$ is the ``in flow,'' i.e. $X^+(v;e)=\max(0,X|_e(v)\cdot n_{e,v})$,
and
\item for any $v$ we have
$$
\sum_{e=\{u,v\}\in E} X^-(v;e) = \left\{ \begin{array}{ll}
	1+c & \mbox{if $v\in A$} \\ 0 & \mbox{otherwise,} \end{array}\right.
$$
where $X^-$ is the ``out flow,'' $X^-(v;e)=\max(0,-X|_e(v)\cdot n_{e,v})
=X^+(u;e)$.
\end{enumerate}
Parts 1--3 of the theorem are now clear.  By
equations~\ref{eq:rho} and \ref{eq:needed_for_alon} we have
$$
\rho_{\sup}(\cg^{|X|^2}) = \sup_v \cv^{-1}(v)\sum_{e\ni v}\ce(e)|X(e)|^2/2
=  \sup_v \sum_{e\ni v}\ell_e |X(e)|^2/2.
$$
Part 4 now follows from the fact
if $x_i\in\reals$ with $0\le x_i\le 1$ and their sum is $d$, then their
sum of squares is at most $\lfloor d\rfloor + (d-\lfloor d\rfloor)^2$.
\proofbox

In lower bounding ${\cal R}(f)$ we may assume that $f\ge 0$, since
we easily see ${\cal R}(|f|)\le {\cal R}(f)$.
Now given $f\in C^1_\Dirichlet(\cg)$ with $f\ge 0$, choose $A$ to be
the support of $f$ and let $X$ be the vector field of Alon described above.
We have
$$
\cq_1(f)\ge \intv{-(\nabla\cdot X)f^2}\biggm/\intv{f^2} \ge c,
$$
and, from equation~\ref{eq:needed_for_alon}
$$
\cq_2(f)\le \sqrt{\rho_{\sup}(\cg^{|X|^2})} \le \sqrt{
[2+\lfloor c\rfloor + (c-\lfloor c\rfloor)^2]/2}.
$$
It follows that:

\begin{theorem}[Alon] For a $c$-magnifier as above, with 
$\cv(v)=a_e=\ell_e=1$ and for all $v,e$ we have that
$$
\lambda \ge c^2/(2c^2+4),
$$
and furthermore, what is more precise for $c>1$,
$$
\lambda \ge c^2\Bigm/
\Bigl(4+2\lfloor c\rfloor + 2(c-\lfloor c\rfloor)^2\Bigr).
$$
If not all $\ell_e$'s are one, the same result holds where we divide the
right-hand-side by $\sup_e\ell_e$.
\end{theorem}

We finish this subsection by showing one way to apply Alon's technique
to the case were the vertex measure is not the counting measure.  So
assume $a_e=1$, but $\cv(v)$ and $\ell_e$ are arbitrary.
First, modify the notion of ``magnifier'' to mean that
$$
\cv\Bigl(\Gamma(A)\Bigr) \ge (1+c)\cv(A)
$$
for all $A\subset V\setminus\partial\cg$.
For such a magnifier and a fixed $A$, 
construct a similar network on $\{s,t\}\cup B_1\cup B_2$, except
(1) the capacity from $s$ to $v$ is $\cv(v)$, (2) the capacity from $v$
to $w$ with $v\in B_1$ and $w\in B_2$ is $\cv(w)$, and (3) the capacity
from $w$ to $t$ is $\cv(w)$.  We get a vector field, $X$, such that
\begin{enumerate}
\item $-\nabla\cdot X$ is $\ge c$ on $A$ and $\le 0$ on $\complement A$,
\item for any $v$ we have $\sum_{e\in v}X^+(v;e)\le\cv(v)$,
\item for any $v$ we have $\sum_{e\in v}X^-(v;e)\le(1+c)\cv(v)$ and each
$X^-(v;e)$ with $e=\{u,v\}$ is less than $\cv(u)$.
\end{enumerate}
Now we can derive various results as before, depending on the values of 
$\cv(v)$ and $\ell_e$.

For example, in \cite{bobkov}, $\ell_e$, where $e=\{u,v\}$ is taken to
be $1/\Bigl( \cv(u)+\cv(v) \Bigr)$.  We conclude
$$
\cv^{-1}(v)\sum_{e\in v}\ce(e)[X^+(v;e)]^2/2 \le 
\cv^{-2}(v)\sum_{e\in v}[X^+(v;e)]^2/2 \le 1/2.
$$
In trying to upper bound the same sum over $X^-$, fix $v$ and view the
$\cv(u)$'s with $e=\{v,u\}$ as variables; given that
$X^-(v;e)\le \cv(u)$ for $e=\{u,v\}$, we can always assume
$X^-(v;e)= \cv(u)$ in maximizing the sum over $X^-$
(or we get a larger sum by taking $\cv(u)$ smaller).
Hence the sum over $X^-$ is bounded by
$$
\cv^{-1}(v)\sum_{e\in v}\ce(e)[X^-(v;e)]^2/2 =
\cv^{-1}(v)(1/2)\sum_{e=\{v,u\}} \frac{\cv^2(u)}{\cv(u)+\cv(v)} =
$$
$$
\cv^{-1}(v)(1/2)\sum_{e=\{v,u\}}\cv(u) - \cv^{-1}(v)(1/2)\sum_{e=\{v,u\}}
\frac{1}{\cv^{-1}(u)+\cv^{-1}(v)}.
$$
Since $1/(1+t)$ is convex, and since the sum of
the $\cv(u)$'s is $\le (1+c)\cv(v)$, it follows that the above sum is
maximized with one $\cv(u)$ equal to $(1+c)\cv(v)$ and the rest zero.
Hence
$$
\cv^{-1}(v)(1/2)\sum_{e\in v}\ce(e)[X^-(v;e)]^2 \le 
\cv^{-1}(v)(1/2) \Bigr(\cv(v)+\cv(v)(1+c)\Bigl)^{-1}(1+c)^2\cv^2(v)
$$
$$
= (1+c)^2/(4+2c).
$$
We conclude:
\begin{proposition} Consider a Rayleigh quotient with $\cv(v)$ arbitrary
and $\ce(e)=\cv(v)+\cv(u)$ for each $e=\{u,v\}$.  Then if $\cg$ is a
$c$-magnifier (in the sense above), then
$$
\lambda\ge c^2(2+c)/(6+6c+2c^2).
$$
\end{proposition}
In \cite{bobkov} a somewhat weaker lower bound is given, although their bound
is based only on a ``part'' of $\lambda_2$, namely what they call
$\lambda_\infty$.

\subsection{Improvements to Cheeger's Inequality for Graphs}

Mohar first gave an improvement to Dodziuk's form of Cheeger's inequality
using the special nature of graphs.  From our point of view there are really
two improvements.  First, Mohar uses
$$
\inte{X\cdot\nabla(f^2)} = 2\inte{\widebar f X\cdot\nabla f},
$$
where $\widebar f$ is the edgewise constant
function whose value on an edge is the average of $f$'s values at the 
endpoints.  Thus we may replace $\cq_2(f)$ with
$$
\wt\cq_2(f)=\|\widebar f\|_{2,\ce}/\|f\|_{2,\cv}.
$$
Next we notice that if $f$ is linear on $e$ 
with $e=\{u,v\}$ and $f(u)=b$ and $f(v)=c$, then
$$
\int_e f^2\,d\ce = a_e(b^2+2bc+c^2)/4,\quad\mbox{and}\quad
\int_e |\nabla f|^2\,d\ce = a_e(b^2-2bc+c^2),
$$
provided that the edge lengths are $1$.  So like equation~\ref{eq:mohar}
we conclude
$$
\|\widebar f\|_{2,\ce}^2 + (1/4)\|\nabla f\|_2^2 = \intv{\rho f^2}\le 
\rho_{\sup}\|f\|_{2,\cv}^2,
$$
with equality if $\cg$ is regular.  We conclude that
$$
\lambda \ge I_\infty^2/[4(\rho_{\sup}-\lambda/4)],
$$
in other words
$$
\lambda^2 - 4\rho_{\sup}\lambda+I_\infty^2\le 0,
$$
i.e.
$$
\lambda \ge 2\rho_{\sup}-\sqrt{4\rho_{\sup}^2-I_\infty^2}.
$$
Notice that for $I_\infty/\rho_{\sup}$ small, this lower bound is within
$O(I_\infty^4/\rho_{\sup}^3)$ of Dodziuk's lower bound of $I_\infty^2/
(4\rho_{\sup})$.  

\ignore{
We claim that it is to be expected that for $I_\infty/\rho_{\sup}$ small,
Mohar's result is, to first order, the same as Dodziuk's.  Namely, if $M$
is a bounded domain in $\reals^n$, we can lay down finer and finer grids to
approximate $M$.  For these grids we have $\rho_{\sup}=n$ and $I_\infty\to
0$ as the grids get finer and finer, and the Rayleigh quotients of the grids
(and therefore $\lambda$) grow proportional to $h^{-2}$ where $h$ is the
grid size.  It follows that Mohar's result were better than a constant factor
times Dodziuk's bound as $I_\infty/\rho_{\sup}\to 0$, 
it would also be an improvement in analysis, for all subdomains of $\reals^n$.
But we know that Mohar's method uses concavities and differences in functions
that are not present in analysis, and that they
become less significant in grid
approximations as the grids get finer.  Hence we don't expect Mohar's 
result to improve Dodziuk's by more than a second order term
for $I_\infty/\rho_{\sup}$ small.
}

\subsection{Closed graphs}

We remark that the inequalities of Dodziuk, Mohar, and Alon all generalize
to closed graphs, where we take $\lambda$ to be the minimum of the
Rayleigh quotient over all functions whose integral with respect to $\cv$
is zero, and where $I_\infty$ and $c$ are isoperimetric constants for
$\Omega$ with $\cv(\Omega)\le\cv(\cg)/2$.  Indeed, $\lambda$ is the
eigenvalue of an eigenfunction, $f$, orthogonal to the first eigenfunction
(the constant function), and by the nodal region
theory of \cite{Fried2}, $f$'s restriction to either nodal region
is the first eigenfunction
of that nodal region.  We can apply any previous result of this subsection
to the nodal region with the smaller $\cv$ measure, to conclude the
analogous theorem for a closed graph.

\section{The Laplacian and the Heat Kernel}\label{se:heat}

One can study the Laplacian by way of the heat equation.  A basic tool
in understanding the heat equation is the (or a) heat kernel, which we define
here.
We are only interested in the ``minimal non-negative'' heat kernel.
However, spectral theory can be used to construct a heat kernel that
comes with many nice bounds; these bounds also apply to the minimal
non-negative heat kernel.  So we shall first construct the ``spectral theory''
heat kernel.  We shall show that the two aforementioned heat kernels are
the same in many interesting cases; we don't know if they are always the
same.

\ignore{
Let $\cg$ admit an orthonormal basis in $L^2_\Dirichlet(\cg,\cv)$, 
$\{\phi_i\}$, of eigenfunctions for
$\Delta$, i.e. $\Delta\phi_i=\lambda_i\phi_i$.  Then
$$
K(x,y,t) = \sum_i e^{-t\lambda}\phi_i(x)\phi_i(y)
$$
is a ``heat kernel,'' in the following sense:
for every $f\in L^2_\Dirichlet(\cg,\cv)$ we
have
$$
u(x,t) = \int K(x,y,t)f(y)\,d\cv(y) 
$$
for $t\ge 0$ is edgewise linear,
satisfies a ``heat equation'' $u_t=-\Delta u$ at each vertex, and satisfies
the ``initial condition'' $u(x,0)=f(x)$.

The goal of this section is to construct the ``heat kernel'' for all
locally finite graphs; this procedure is well-known (see \cite{davies}).
Notice that if $z\in\interior V$, then the function
$\delta_z=\chi_z/\cv(z)$ is an analogue of the Dirichlet delta function at
$z$ in that
$$
\int f(y)\delta_z(y)\,d\cv(y) = f(z).
$$
It follows that if $K(x,y,t)$ is a heat kernel in the above sense,
then for any fixed $z$, $K(x,z,t)$ is a solution to the heat equation as
above with initial condition $K(x,z,0)=\delta_z(x)$.

It follows that one can try to construct the heat kernel $K(x,y,t)$ by
looking for the solution to the heat equation with initial condition
$\delta_y(x)$.  Unfortunately, the solution to the heat equation is not
generally unique; we point this out in the next subsection (but this
subsection has no further use in this paper).  The usual appoach to 
constructing $K(x,y,t)$ is to pick out a ``preferred'' solution to
the heat equations above, namely $e^{-t\Delta_x}\delta_y(x)$, where
$\Delta_x$ means the Laplacian in the $x$ variable, and where the
aforementioned exponential has a reasonable intepretation.  First we
will interpret the ``heat operator'' $e^{-t\Delta}$ and give some of
its properties, then we will discuss the resulting properties of
$K(x,y,t)$.  We emphasize that we primarily discuss only those properties
of $K(x,y,t)$ needed for our later discussion of the Nash estimates.
}

\subsection{The Dirichlet Initial Value Problem and the Heat Kernel}

The heat kernel is intuitively obtained by solving many instances of
the ``heat equation.''  When the heat equation has a unique solution,
then there is a unique heat kernel.  When the heat equation solution is not
unique, there may be more than one heat kernel.
Here we describe what is meant by solving the heat equation, and
illustrate cases where the heat equation
does not have a unique solution.

\begin{definition} A
function $u=u(x,t)\from\cg\times[0,T]\to\reals$ for some
$T>0$ is said to satisfy the {\em heat equation} on $\cg\times[0,T]$ if
(1) $u$ is continuous on $\cg\times[0,T]$,
(2) $u(\;\cdot\;,t)$ is edgewise linear for each $t$,
(3) we have $u_t$ (the partial derivative of $u$ with respect to $t$)
and $-\Delta u$ (the Laplacian in the variable $x$) exist at the point
$(x,t)$ and are equal there, for all
$x\in \interior V$ and $t\in(0,T)$.  We can also take $T=\infty$ in the
above, replacing $[0,T]$ with $[0,\infty)$.
\end{definition}

Given a function $f$, on $\interior V$, we seek a solution 
to the heat equation, $u(x,t)$ satisfying the two ``boundary conditions''
$$
u(x,0)=f(x) \quad\forall x\in \interior V
$$
and
$$
u(x,t)=0 \quad\forall x\in\partial\cg,\;\forall t\in[0,T].
$$
We will call this the ``Dirichlet initial value problem'' for the heat
equation (with ``initial value'' $f$).



The following theorems give examples of
non-uniqueness or uniqueness of the heat kernel, and we will prove them in appendix B
(since they are not crucial to 
our later use of the heat kernel).

\begin{theorem}\label{th:finite_uniqueness}
 If $\cg$ is finite, then any Dirichlet initial value problem
has at most (in fact, exactly) one solution.
\end{theorem}

By way of contrast we have non-uniqueness for a very ``mildly'' infinite
graph.  This example is
a simple adaptation of the example in \cite{friedman-parabolic},
page 31, to graphs.

\begin{theorem}\label{th:mild_graph_nonuniqueness}
Let $\cg$ be the graph whose vertices are the integers,
$\integers$, with one edge from $i$ to $i+1$ for all $i\in\integers$.
Then the Dirichlet initial value problem with initial value $0$ has
infinitely many solutions.
\end{theorem}

However, the infinitely many solutions referred to above are unbounded
even for fixed $t$.  So one might hope for a unique solution to the heat
equation that is bounded over $\cg$ for any fixed time, $t$.

\begin{theorem}\label{th:growing_degree_uniqueness} 
For every $v \in \interior V$, set 
$L(v)= \cv(v)^{-1} \sum_{e \ni v} a_e/l_e$ and
let $L_i(v)$ denote the supremum of $L(u)$ over all $u$
with a path to $v$ through $\interior V$ of length at most $i$.  If
there is a $C$ such that
for any fixed $v\in\interior V$ we have
$L_j(v)\le Cj$ for sufficiently large $j$, then the
Dirichlet initial value problem has at most (in fact, exactly) one 
solution, $u(x,t)$, bounded in $x$ for any fixed $t$.  
The same is true if there is a $C$ such that
for any fixed $v\in\interior V$ 
we have 
$$
L_0(v)L_1(v)\cdots L_j(v)\le (Cj)^j
$$
for sufficiently large $j$.
\end{theorem}

However, this theorem is close to being the best possible, in a sense:

\begin{theorem}\label{th:growing_degree_nonuniqueness}
For any $\alpha>0$ there exists a tree, $\cg$, such that for any fixed
$v\in\interior V$ we have
$L_j(v)\le 2j^{1+\alpha}$ for sufficiently large $j$, and such that
the Dirichlet initial value problem has infinitely many solutions that are
bounded on $\cg\times[0,T]$ for any fixed $T$.
\end{theorem}

Our last theorem looks like a precursor to a uniqueness theorem.  However,
we will use it in establishing positivity for the heat kernel, 
so we give its simple proof here.

\begin{theorem}\label{th:simple_maximum_principle}
Let $u\ge0$ be a solution to the heat equation over $[0,T]$.  
Let
$u(y,t_0)=0$ with $0<t_0\le T$ and $y\in\interior V$.  Then $u(x,t_0)=0$
for any $x$ that is connected to $y$ via a (finite) path of vertices in
$\interior V$.
\end{theorem}
\proof Since $u\ge0$ we have $u_t(y,t_0)\le 0$ and so $\Delta u(y,t_0)\ge 0$.
Hence
$$
0 =\sum_{e\sim\{y,w\}} (a_e/\ell_e) u(y,t_0) \ge
\sum_{e\sim\{y,w\}} (a_e/\ell_e) u(w,t_0),
$$
and we conclude that $u(w,t_0)=0$ for all $w\in\interior V$ with an edge to
$y$.  Repeatedly applying this conclusion yields the theorem.
\proofbox

\subsection{Heat Kernels}

Let $\delta_v=\chi_v/\cv(v)$; $\delta_v$ is an analogue of the Dirac delta
function at $v$.
\begin{definition} A {\em fundamental solution}, 
$K(x,y,t)$, for the heat equation
is a function $K\from\cg\times\cg\times\reals_{\ge 0}\to\reals$ (where
$\reals_{\ge 0}$ are the non-negative reals) such that 
(1) for any fixed $y\in\interior V$
we have $K(x,y,t)$, viewed as a function of $x,t$, is a solution to the
Dirichlet initial value problem with initial value $\delta_y$, (2) for any
$y\in\partial\cg$ we have $K(x,y,t)=0$, and (3) $K(x,y,t)$ is edgewise linear
in $x$ and $y$.
\end{definition}
Conditions (2) and (3) imply that a fundamental solution is determined by
solving the Dirichlet initial value problem at all interior vertices.

\begin{definition} A fundamental solution, $K(x,y,t)$, as above is said to
be a {\em heat kernel} if (1) it is symmetric, i.e. $K(x,y,t)=K(y,x,t)$,
(2) it is self-reproducing, i.e. 
\begin{equation}\label{eq:self-reproducing}
K(x,y,t) = \int K(x,s,\tau)K(s,y,t-\tau)\,d\cv(s)
\end{equation}
for any $0\le\tau\le t$, and (3) we may formally differentiate 
equation~\ref{eq:self-reproducing} in $t$, i.e.
$$
K_t(x,y,t) = \int K(x,s,\tau)K_t(s,y,t-\tau)\,d\cv(s).
$$
\end{definition}
Condition (3) is not usually stated in the definition of a heat kernel, 
but this condition will be needed in Nash's technique
in section 7; it usually holds if (2) holds.

\begin{definition} A heat kernel, $K$, is {\em non-negative} if $K(x,y,t)\ge 0$
for all $x,y,t$.  If there is a heat kernel,
$K$, such that $K\le G$ (pointwise) for any
non-negative fundamental solution, $G$, to the heat equation, we say that
$K$ is the minimal non-negative heat kernel.
\end{definition}
We shall show that for locally finite graphs there always exists a minimal
non-negative heat kernel.  We begin by giving a natural and well-known
construction of a heat
kernel using spectral theory; this heat kernel is not always the minimal
non-negative one, but many important bounds hold for the spectral theory
heat kernel, and these bounds immediately follow for the minimal heat kernel
as well.

Theorem~\ref{th:simple_maximum_principle} shows that if $K$ is a
non-negative fundamental solution and $x$ and $y$ are connected in
$\interior\cg$, then $K(x,y,t)$ is strictly positive for all $t>0$.

\subsection{The Heat Operator}

In this section we give meaning to the expression $e^{-t\Delta}$, which
we call the heat operator.  We do so using the theory of quadratic forms
and unbounded self-adjoint operators as in \cite{davies,davies-semigroups}.

Let $\cd$ denote the subspace of $L^2_\Dirichlet(\cg,\cv)$ consisting of
functions of finite type; notice that $\cd$ is dense there.
Let $H^1(\cg)$ denote the closure of $\cd$ under the norm
$$
\|f\|_{H^1}^2= (f,f) + (\nabla f,\nabla f).
$$
It is easy to see that
the quadratic form $Q(f,g)=(\Delta f,g)=(\nabla f,\nabla g)$ 
defined on $\cd$ (i.e. defined when $f,g\in\cd$)
extends to an quadratic form, $\closure Q$, on $H^1(\cg)$, 
known as the {\em closure} of $Q$.  We identify $\closure Q$ with $Q$ when
no confusion can arise.
It is easy to check that we may view $H^1(\cg)$ as a subspace
of $L^2_\Dirichlet(\cg,\cv)$ (see the bottom of page 106 in
\cite{davies-semigroups}).

\begin{proposition}\label{pr:invariant}
If $f\in H^1(\cg)$ then $|f|\in H^1(\cg)$ and
$Q(|f|,|f|)\le Q(f,f)$.  The same is true if $|f|$ is replaced by
$g=\max\Bigl( 0,\min(f,1)\Bigr)$.
\end{proposition}
\proof  First we remark that if $h\in\cd$ then it is easy to check that
$|h|\in\cd$ and $Q(|h|,|h|)\le Q(h,h)$.
If $f\in H^1(\cg)$ then there exist $f_n\in\cd$ with $\|f-f_n\|_{H^1}
\to 0$.  Then $|f_n|\in\cd$, and we easily see that
$$
\|\;|f|-|f_n|\;\|_{H^1}\le \|f-f_n\|_{H^1}\to 0
$$
so that $|f|\in H^1(\cg)$.  $Q(|f_n|,|f_n|)\le Q(f_n,f_n)$ now shows
$Q(|f|,|f|)\le Q(f,f)$.
The proof of the statement in the proposition
involving $g$ is the same.
\proofbox

Now we invoke some spectral and quadratic form theory.  According to
theorems 4.12 and 4.14 in \cite{davies-semigroups} and their proofs, 
$\Delta$ can be extended to 
self-adjoint operator, which we again call $\Delta$, whose domain includes
$\cd$ (and lies in $H^1(\cg)$); furthermore $H^1(\cg)$ equals ${\rm Quad}
(\Delta)$.  By the spectral theorem 4.4 in
\cite{davies-semigroups}, it makes sense to speak 
of $e^{-t\Delta}$ for
any $t\ge 0$ as a bounded operator on the closure of $\cd$ in
$L^2_\Dirichlet(\cg,\cv)$, which is just $L^2_\Dirichlet(\cg,\cv)$.  
We call $e^{-t\Delta}$ the ``heat operator.''

Using the spectral theorem it is easy to see
that $e^{-\Delta t}$ has norm at most $1$,
and that $e^{-t\Delta}$ is strongly continuous, i.e. that
$e^{-\Delta t}f$ is $L^2$ continuous in $t$ for any 
$f\in L^2_\Dirichlet(\cg,\cv)$.

The proof of lemma 1.3.4 (and theorems 1.3.2 and 1.3.3)
in \cite{davies} 
show that
proposition~\ref{pr:invariant} implies that $e^{-\Delta t}$ is
\begin{enumerate}
\item {\em positivity preserving}, meaning that if $f\in
L^2_\Dirichlet(\cg,\cv)$
satisfies $f\ge 0$ everywhere, then so does $e^{-\Delta t}f$, and
\item {\em a contraction} on $L^p_\Dirichlet(\cg,\cv)$ 
for any $p\in[1,\infty]$, meaning that if
$f\in L^2_\Dirichlet(\cg,\cv)\cap L^p(\cg,\cv)$ then the same is true of
$e^{-\Delta t}f$ and we have $\|e^{-\Delta t}f\|_p
\le \|f\|_p$.
\end{enumerate}

\subsection{The Spectral Theory Heat Kernel}

For $v\in\interior V$ let $\delta_v$ be the edgewise linear function that
is $1/\cv(v)$ on $v$ and $0$ on other vertices.
$\delta_v$ is the analogue of the ``Dirichlet delta function'' in graph 
theory.

\begin{definition} The {\em spectral theory heat kernel} is the function
$\wt K\from \interior V\times\interior V\times[0,\infty)\to\reals$ defined by
$$
\wt K(x,y,t) = (e^{-t\Delta}\delta_y,\delta_x) = (e^{-t\Delta}\delta_y)(x).
$$
(Since $e^{-t\Delta}$ is a bounded operator defined on all
of $L^2_\Dirichlet(\cg,\cv)$ and 
$\delta_y$ lies there, this definition makes sense.)
\end{definition}

Keeping the notation of the last subsection, the following proposition
is easy.

\begin{proposition} $\wt K(x,y,0)=\delta_y(x)$.  For fixed $y,t$ we have that
$\wt K(\,\cdot\, ,y,t)$ lies in $L^2_\Dirichlet(\cg,\cv)$, and for any
$f\in L^2_\Dirichlet(\cg,\cv)$ and $x$ we have
$$
(e^{-t\Delta}f)(x) = \int \wt K(x,y,t)f(y)\,d\cv(y).
$$
Finally
$\wt K$ is symmetric and self-reproducing.  In other
words, $\wt K(x,y,t)=\wt K(y,x,t)$ and
$$
\wt K(x,y,t) = \int \wt K(x,s,\tau)\wt K(s,y,t-\tau)\,d\cv(s)
$$
for any $0\le\tau\le t$.  
\end{proposition}

\proof The first claim is clear.  The symmetry of $\wt K$ follows from the
self-adjointness of $e^{-t\Delta}$, i.e.
$$
\wt K(x,y,t) = (e^{-t\Delta}\delta_y,\delta_x) = (\delta_y,e^{-t\Delta}\delta_x)
=\wt K(y,x,t).
$$
Since $e^{-t\Delta}$ contracts the $L^2$ norm, and since $\delta_y\in
L^2_\Dirichlet(\cg,\cv)$ for any $y\in\interior V$, we have 
$e^{-t\Delta}\delta_y\in L^2_\Dirichlet(\cg,\cv)$ for any $t\ge 0$, and
hence the claim about $K(\,\cdot\, ,y,t)$ being in $L^2$.  For any
$f\in L^2_\Dirichlet(\cg,\cv)$ and $x$ we have
$$
\int \wt K(x,s,t)f(s)\,d\cv(s)=\int (e^{-t\Delta}\delta_x)(s)f(s)\,d\cv(s)
$$
$$
=(e^{-t\Delta}\delta_x,f) = (\delta_x,e^{-t\Delta}f)= (e^{-t\Delta}f)(x).
$$
For the last claim we apply the last formula with $f(s)=\wt K(s,y,t-\tau)$ to
conclude
$$
\int \wt K(x,s,\tau)\wt K(s,y,t-\tau)\,d\cv(s)=
\Bigl( e^{-\tau\Delta}\wt K(\,\cdot\, ,y,t-\tau)\Bigr)(x) 
$$
$$
= (e^{-\tau\Delta}e^{-(t-\tau)\Delta}\delta_y)(x) =
(e^{-t\Delta}\delta_y)(x) = \wt K(x,y,t).
$$
\proofbox

\begin{proposition} For all $x,y\in\interior V$ and $t>0$ we have:
$$
\Delta_x \wt K(x,y,t) = - \wt K_t(x,y,t) = (\Delta e^{-t\Delta}\delta_x,\delta_y)
$$
where $\Delta_x$ denotes the Laplacian in the $x$ variable.
\end{proposition}
Similarly $\wt K_{tt}(x,y,t)$ exists and satisfies a similar formula as above;
the proof below easily generalizes.
\proof The spectral theorem and Taylor's theorem easily imply that the
operator $e^{-t\Delta}$ is differentiable, with limits of Newton quotients
taken with respect to the $L^2$ operator norm, and has derivative
$\Delta e^{-t\Delta}$.  It follows that
$$
-\wt K_t(x,y,t)=-\frac{\partial}{\partial t}(e^{-t\Delta}\delta_x,\delta_y)
= (\Delta e^{-t\Delta}\delta_x,\delta_y).
$$
Also
$$
\Delta_x \wt K(x,y,t) = \Delta_x (e^{-t\Delta}\delta_y)(x) =
(\Delta e^{-t\Delta}\delta_y)(x) = (\Delta e^{-t\Delta}\delta_y,\delta_x).
$$
\proofbox

The above proposition and the same type of calculation used to show that
$\wt K$ is self-reproducing also shows:
\begin{proposition} For any $x,y\in\interior V$ and $0\le\tau\le t$ we have
$$
\wt K_t(x,y,t) = \int \wt K(x,s,\tau)\wt K_t(s,y,t-\tau)\,d\cv(s).
$$
In other words, the self-reproducing property of $\wt K$ can be partially
differentiated with respect to time (i.e. $t$) within the integration.
\end{proposition}

Now we outline a proof of the following result, needed here only for
subsection 6.1.

\begin{theorem} Let $f\in L^\infty_\Dirichlet(\cg,\cv)$.  Then
$$
u(x,t)=\int \wt K(x,y,t)f(y)\,d\cv(y)
$$
solves the Dirichlet initial value problem with initial value $f$
(the $L^\infty$
contractive property of $e^{-t\Delta}$ shows that $\wt K(x,\,\cdot\, ,t)$
is in $L^1$, and so the above integral exists).
\end{theorem}
We remark that the same theorem is true with $L^\infty$ replaced by
$L^p$ for any $1\le p\le\infty$, and all aspects of the proof below are
the same or easier.
\proof 
Clearly $u(x,0)=f(x)$.

We may assume $\cg$ is connected, and since it is locally finite
we can enumerate its vertices $V=\{v_1,v_2,\ldots\}$.  For each $n$
consider the function $f_n$ which agrees with $f$ on $v_1,\ldots,v_n$ and
is zero on other vertices.  Set
$$
u_n(x,t)=\int \wt K(x,y,t)f_n(y)\,d\cv(y).
$$
Since the above integral represents a finite sum, we see that each $u_n$
satisfies the heat equation.

Note that $|f| \wt K(x,\,\cdot\, ,t)=|f| e^{-t\Delta}\delta_x$
is in $L^1$ and  from the Lebesgue dominated convergence theorem it follows that for fixed $x$,
$u_n(x,t)\to u(x,t)$ as $n\to\infty$.

From now on, we fix $x\in \interior V$ and $t>0$.

We first show that $u(x,t)$ is continuous in $t$. 
It suffices to notice that
\begin{equation}
\label{eq:continuity}
u_n(x,t+h)-u_n(x,t) = h \frac{\partial u_n}{\partial t} (x,t_n)
\end{equation}
for some $t_n$ between $t$ and $t+h$.

The right-hand-side term can be bounded by noting that
$$
(u_n)_t(x,s) = (-\Delta e^{-s\Delta}f_n,\delta_x) =
(e^{-s\Delta}f_n,-\Delta\delta_x).
$$
and
$$
|(u_n)_t(x,s)|\le \|e^{-s\Delta}f_n\|_\infty \|\Delta \delta_x\|_1
\le \|f\|_\infty \|\Delta \delta_x\|_1.
$$
$\cg$ is locally finite, therefore $\|\Delta^2\delta_x\|_1$
is finite.  Taking $n\to\infty$ in equation~\ref{eq:continuity} and
using the pointwise convergence of $u_n$ we conclude
$$
|u(x,t+h)-u(x,t)| \le h
\|f\|_\infty \|\Delta\delta_x\|_1.
$$
Taking $h\to 0$ we conclude that $u(x,t)$ is continuous in $t$.

It can be shown that $\Delta u = -u_t$ by a similar proof technique, where we  
start by noting that
\begin{equation}\label{eq:close_to_heat}
\Delta u_n = -(u_n)_t = -\frac{u_n(x,t+h)-u_n(x,t)}{h} + (h/2)(u_n)_{tt}(
x,\wt t_n)
\end{equation}
for some $t_n$ between $t$ and $t+h$.
\proofbox
\ignore{  Since $\wt K(x,\,\cdot\, ,t)=e^{-t\Delta}\delta_x$
is in $L^1$, we have for fixed $x$ and $t$ and any $\epsilon>0$
there is a finite $A\subset
\interior V$ such that
$$
\int_{\complement A} \wt K(x,y,t)\,d\cv(y) = (\chi_{\complement A},
e^{-t\Delta}\delta_x) < \epsilon/2.
$$
Since $\chi_{\complement A},\delta_x\in L^2$ and $e^{-t\Delta}$ is strongly
continuous, it follows that there is an $\eta>0$ such that
$$
\int_{\complement A} \wt K(x,y,\wt t)\,d\cv(y) = (\chi_{\complement A},
e^{-\wt t\Delta}\delta_x) < \epsilon
$$
for all $\wt t$ with $|\wt t - t|<\eta$.  It follows that for fixed $x$
$u_n(x,t)\to u(x,t)$ as $n\to\infty$ and that $u(x,t)$ is continuous in $t$.
Also, clearly $u(x,0)=f(x)$.

It remains to show that $\Delta u = -u_t$.  Fix
$x\in \interior V$, $t>0$, and $h$ small, and notice that
\begin{equation}\label{eq:close_to_heat}
\Delta u_n = -(u_n)_t = -\frac{u_n(x,t+h)-u_n(x,t)}{h} + (h/2)(u_n)_{tt}(
x,\wt t_n)
\end{equation}
for some $\wt t_n$ with $t<\wt t_n<t+h$.  Now
$$
(u_n)_{tt}(x,s) = (\Delta^2 e^{-s\Delta}f_n,\delta_x) =
(e^{-s\Delta}f_n,\Delta^2\delta_x).
$$
It follows that
$$
|(u_n)_{tt}(x,s)|\le \|e^{-s\Delta}f_n\|_\infty \|\Delta^2\delta_x\|_1
\le \|f\|_\infty \|\Delta^2\delta_x\|_1.
$$
Since $\cg$ is locally finite, it follows that $\|\Delta^2\delta_x\|_1$
is finite.  Taking $n\to\infty$ in equation~\ref{eq:close_to_heat} and
using the pointwise convergence of $u_n$ and the local finiteness of
$\cg$ we conclude
$$
\left| \Delta u + \frac{u(x,t+h)-u(x,t)}{h} \right| \le \frac{h}{2}
\|f\|_\infty \|\Delta^2\delta_x\|_1.
$$
Taking $h\to 0$ we conclude that $u_t$ and equals $-\Delta u$.
}

\subsection{The Miminal Heat Kernel}\label{sb:finite_limit}

The minimal non-negative heat kernel, $K$, of a locally finite graph,
$\cg$, can be constructed as 
the limit of the heat kernels of any sequence of finite
graphs that ``exhaust'' $\cg$.  We prove this here and give the properties
that follow.  Bounds on $\wt K$ of the last subsection will apply to $K$.
We will show that $\wt K=K$ in a number of interesting cases, but we don't
know if this is generally true.

As usual, let $\interior V$ be the set of interior vertices of $\cg$.
For $A\subset \interior V$, 
let $K_A=K_A(x,y,t)$ be the spectral heat kernel on the graph $\cg_A$,
which is the same as $\cg$ except that the set of boundary vertices is
all vertices not in $A$\footnote{
	In particular, $K(x,y,t)$ is defined for all $x,y\in\cg$, identifying
	the geometric realization of $\cg$ with that of $\cg_A$.  Also
	$K_A(x,y,t)$ vanishes if $x$ lies on an edge whose endpoints don't
	lie in $A$.
}.  Since $\cg_A$ has only finitely many interior vertices and edges (an
interior edge is one with at least one endpoint in the interior), there
is a finite orthonormal basis of eigenfunctions for 
$L^2_\Dirichlet(\cg_A,\cv)$, $\phi_1,\phi_2,\ldots$, with corresponding
eigenvalues $\lambda_1,\lambda_2,\ldots$, and we clearly have
$$
K_A(x,y,t)=(e^{-t\Delta_A}\delta_x,\delta_y) = \sum_i e^{-t\lambda_i}
\phi_i(x)\phi_i(y),
$$
where $\Delta_A$ is the Laplacian associated to $\cg_A$; more explicitly,
$\Delta_A=\pi_A\Delta\pi_A$, where $\Delta$ is the Laplacian on $\cg$ and
$\pi_A$ is the projection on functions on $\cg$ that sends $f$ to
$0$ on boundary vertices and $f(v)$ on interior vertices $v$ (and is extended
to be edgewise linear).

We remark that when no confusion can arise, we will sometimes drop the
$A$ from $\Delta_A$.

Consider any sequence $A_1,A_2,\ldots$ such that
(1) $A_i\subset\interior V$, (2) each $A_i$ is finite, (3) $A_i\subset 
A_{i+1}$, and (4) for all $v\in\interior V$ there is an $i$ with 
$v\in A_i$.  We will say that the sequence $A_i$ is a {\em increasing
finite set exhaustion of $\cg$}.

\begin{theorem} For any increasing finite set exhaustion, $A_i$, of $\cg$,
and any fixed $x,y,t$ we have $K_{A_i}(x,y,t)$ is non-decreasing
in $i$, tending to a limit $K(x,y,t)$.  $K$ is independent of the set
exhaustion, and $K\le G$ for any non-negative fundamental solution to the
heat equation.
\end{theorem} 

\proof  We shall need the following very easy maximum principle.

\begin{lemma} Let $u,v$ be two solutions to the heat equation in
$\cg\times[0,T]$ such that $u\le v$ on the ``boundary,'' namely
$\cg\times\{0\}$ union $\partial\cg\times[0,T]$.  If $\cg$ is finite then
$u\le v$ on all of $\cg\times [0,T]$.
\end{lemma}

\proof If $u>v$ at $(x,t)$
then $u(x,t)>v(x,t)+\epsilon t$ for some positive $\epsilon>0$.  Since
$\cg$ is finite, $w(x,t)=u(x,t)-v(x,t)-t\epsilon$ attains a
maximum, $M>0$, over $\cg\times[0,T]$ somewhere, say at $(x_0,t_0)$.
However, $\Delta w + w_t = -\epsilon <0$, and it follows that either
$w_t$ or $\Delta w$ is negative at $(x_0,t_0)$.  Since $t_0>0$ and $w$
is maximized at $(x_0,t_0)$, $w_t$ cannot be negative there.  But
the same argument
as in theorem~\ref{th:simple_maximum_principle} shows that if $\Delta w$
is negative there, then $w(x,t_0)$ is greater than $w(x_0,t_0)$ for some
neighbour, $x$, of $x_0$, which is again impossible.
\proofbox

Continuing with the proof of the theorem, the maximum principle applied to
$\cg_{A_i}$ shows that
for any $x,y,t$ we have $K_{A_i}(x,y,t)$ is non-decreasing in $i$ and
is $\le G(x,y,t)$ for any non-negative
fundamental solution to the heat equation.  Also the maximum principle
shows $K_A(x,y,t)\le K_B(x,y,t)$
provided that $A\subset B$, which completes the proof.
\proofbox

\begin{proposition} $K$ constructed above is a heat kernel.
\end{proposition}
\proof
Clearly $K(\,\cdot\, ,y,0)=\delta_y$.  We wish to show that $K$ is
continuous in $t$ for $t\ge 0$ and differentiable in $t>0$ and
satisfies the heat equation there.  We integrate the heat equation that
$K_{A_i}$ satisfies (for $i$ sufficiently large as a function of $x$) 
to conclude
$$
-\int_0^t \Delta_x K_{A_i}(x,y,s) = K_{A_i}(x,y,t)-K_{A_i}(x,y,0).
$$
Clearly we can pass this equation to the $i\to\infty$ limit to conclude
the same for $K$ instead of $K_{A_i}$.  $L^\infty$ bounds on $\wt K$
imply the same for $K$, and we first conclude that $K$ is continuous in
time, and then differentiable for $t>0$ and satisfying the heat equation
there.

Since the $K_{A_i}$ are symmetric, so is $K_A$.  Similarly we conclude
that $K$ is self-reproducing, using the bounded convergence theorem;
indeed, the integrand in
\begin{equation}\label{eq:K_A_i_self-reproducing}
\int K_{A_i}(x,s,\tau)K_{A_i}(s,y,t-\tau)\,d\cv(s)
\end{equation}
is non-negative and bounded by
$$
\wt K(x,s,\tau)\wt K(s,y,t-\tau)\,d\cv(s),
$$
whose integral is bounded.  Hence the $i\to\infty$ limit of 
equation~\ref{eq:K_A_i_self-reproducing} is the integral with $K$
replacing $K_{A_i}$.  It easily follows that $K$ is self-reproducing.
Finally, to see that the self-reproducing equation can be differentiated in
$t$, we write
$$
(K_{A_i})_t(x,y,t)=-\Delta_y K_{A_i}(x,y,t),
$$
and
$$
\int K_{A_i}(x,s,\tau)(K_{A_i})_t(s,y,t-\tau)\,d\cv(s) =
\int K_{A_i}(x,s,\tau)\Delta_y K_{A_i}(s,y,t-\tau)\,d\cv(s).
$$
We know that the two left-hand-sides are equal.  The first left-hand-side
clearly tends to $-\Delta_y K(x,y,t)$ as $i\to\infty$.  For the second
left-hand-side, we use the bounded convergence theorem, using an $L^1$ bound
on $K_{A_i}(x,s,t)$ (coming from $\wt K$) and an $L^\infty$ bound on
$\Delta_y  K_{A_i}(s,y,t-\tau)$ via
$$
|\Delta_y  K_{A_i}(s,y,t-\tau)| = |(\delta_s,e^{-(t-\tau)\Delta}\Delta
\delta_y)| \le \|e^{-(t-\tau)\Delta}\Delta\delta_y\|_\infty \le
\|\Delta\delta_y\|_\infty,
$$
the last quantity being bounded for fixed $y$.  We conclude that
$$
-\Delta K(x,y,t) = \int K(x,s,\tau)\Delta_y K(s,y,t-\tau)\,d\cv(s).
$$
Hence the self-reproducing equation can be differentiated in $t$, and
hence $K$ is a heat kernel.
\proofbox

Next we ask when is $K=\wt K$.  An easy proposition is: 
\begin{proposition} Let the Dirichlet initial value problem always have at
most one solution $u(x,t)$ for $x\in\cg$ and all $t\ge 0$ that is uniformly
bounded.  Then $K=\wt K$.
\end{proposition}
Notice that the hypothesis of this proposition is satisfied, 
according to theorem~\ref{th:growing_degree_uniqueness}, 
if $\Delta$ is bounded, or
more generally $L_i(v)$ grows linearly in $i$ for $v$ fixed.
\proof
For $y$ fixed, $\wt K(x,y,t)$ is a solution to a Dirichlet initial value
problem, bounded (by $1/\cv(y)$)
in $x$ and $t$ by the
$L^\infty$ contractivity operator $e^{-t\Delta}$.  $K(x,y,t)$ solves the
same Dirichlet initial value problem, and the same bound holds for
$K(x,y,t)$, since $0\le K\le \wt K$.
\proofbox

We give a different criterion for $K$ to equal $\wt K$.
\begin{proposition} Let $\cv(v)\ge C$ for some constant $C>0$.
Then $K=\wt K$.
\end{proposition}

\ignore{
Under the first assumption we have $L(v)\le C$ for some constant $C$.
Let $\pi_i$ be $\pi_{A_i}$ as in the beginning of this subsection, so that
$\Delta_i=\pi_i\Delta\pi_i$ is the $\cg_{A_i}$ Laplacian extended 
appropriately by $0$ to $\cg$.  $L(v)\le C$ shows that $\|\Delta_i\|\le 2C$.
It follows that
$$
\| e^{-t\Delta_i f} - (I-t\Delta_i+\cdots+(-t\Delta_i)^k/k!)f \| \le
|2Ct|^{k+1}\|f\|/(k+1)!
$$
$$
check this!
$$
and similarly with $\Delta$ replacing $\Delta_i$
We have $\Delta_i^k\to\Delta^k$ strongly for any fixed $k$, i.e. for any
$f\in L^2_\Dirichlet(\cg,\cv)$, $\Delta_i^k f\to\Delta^k f$.  
It follows that
$$
\limsup \| e^{-t\Delta_i}f-e^{-t\Delta}f\| \le 2|2Ct|^{k+1}\|f\|/(k+1)!
$$
for any $k>0$.  Hence $e^{-t\Delta_i}f\to e^{-t\Delta}f$, and it follows that
the heat kernels agree.
\proofbox
}

The maximum principle shows that $\wt K(x,y,t)\le K_{A_i}(x,y,t)+\eta(t,y,A_i)$,
where
$$
\eta(t,y,A) = \sup_{z\in\partial\cg_A,\; 0\le\tau\le t} \wt K(z,y,\tau).
$$
It remains to see that $\eta(t,y,A_i)\to 0$ as $i\to\infty$ for any fixed
$t,y$.  

Fix a value of $y$ and $t$.
The fact that $\eta(t,y,A_i)\to 0$ as $i\to\infty$ follows from the $L^1$
boundedness of $e^{s\Delta}\delta_y$ and its continuity in $s$.
Indeed, for every $\tau_0\in [0,t]$ we have 
$$
\sum_x \wt K(x,y,\tau_0)\cv(x)
$$
is finite.  So for any $\epsilon>0$ there is
some $i_0$ for which 
$$
\sum_{x\notin A_{i_0}} \wt K(x,y,\tau_0)\cv(x) < C\epsilon,
$$
and hence
$\wt K(x,y,\tau_0)<\epsilon$ 
for any $x\notin A_{i_0}$.  By
the  $L^2$ continuity of $e^{s\Delta}\delta_y$ ,
there is an $\alpha>0$ with $\|\wt K(.,y,\tau)- \wt K(.,y,\tau_0)\|_2^2<  \frac{\epsilon^2}{C}$ for all 
$\tau$ with $|\tau-\tau_0|<\alpha$ (and $\tau\ge 0$) and
therefore  
$| \wt K(x,y,\tau)-K(x,y,\tau_0)| < \epsilon$ for all $x$.
Hence $\wt K(x,y,\tau)<
2\epsilon$ for all $x\notin A_{i_0}$ and $ |\tau-\tau_0|<\alpha$.
The topological compactness of $[0,t]$ implies that it can be covered with
a finite set of intervals $\{\tau \;|\;|\tau-\tau_0|<\alpha\}$ as above,
and we conclude $\eta(t,y,A_i)<2\epsilon$ for $i$ greater than the largest
of the finite set of $i_0$'s that correspond to the finite covering 
intervals of 
$[0,t]$.  Since $\epsilon>0$ was arbitrary, we conclude
that $\eta(t,y,A_i)\to 0$ as $i\to\infty$.
\proofbox

In the sections to follow we will use the fact that
if we want to prove an upper bound on $K(x,y,t)$ for fixed $x,y,t$, it
suffices to do so for all $K_A(x,y,t)$ with $A$ finite.

\section{$L^p$ Estimates}

The Federer-Fleming theorems give us lower bounds on the $L^1$ norm
of $\nabla f$.
Other ``gradient estimates,'' namely $L^p$ estimates for $p>1$, i.e.
lower bounds on $\|\nabla f\|_p$, follow readily from the $L^1$ estimates.
Such bounds include Sobolev and Nash inequalities, Trudinger inequalities,
and resulting heat kernel and eigenvalue estimates.

\subsection{$L^p$ Estimates for $p>1$}

It is easy to use the $L^1$ gradient estimates to obtain $L^p$ estimates
in the non-closed situation.  The main new observation here is that often
the analogous inequality in the closed situation follows easily using
``split'' functions; previous works (e.g. \cite{cheng-li,ChuYausob}) use
balanced functions, which makes the estimating more difficult and weaker
by a small multiplicative constant.

\begin{proposition}\label{pr:general}
Let $F\from\reals \to\reals$ be a
differentiable function which preserves sign (i.e. $F(x)$ is positive,
zero, or negative according to whether or not $x$ is).  Let $1\le p\le
\infty$,
and further assume that
$(F')^{p'}$ is convex.  For any $\phi\in C^1_\Dirichlet(\cg)$ we have
$$
I_\nu \| F(\phi) \|_{\nu'} \le \rho_{\sup}^{1/p'} \| \nabla \phi \|_p 
\| F'(\phi)\|_{p'} .
$$
Similarly, in the closed case we have that for any split 
$\phi\in C^1_\Dirichlet(\cg)$ we have the same as above with $\wt I_\nu$
replacing $I_\nu$.
\end{proposition}
\proof The left-hand side comes from applying corollary~\ref{cr:split} or
equation~\ref{eq:estimate} to 
$F(\phi)$, yielding
$$
I_\nu\| F(\phi) \|_{\nu'} \le \Bigl\|\nabla\Bigl( F(\phi) \Bigr)\Bigr\|_1,
$$
and $\wt I_\nu$ replacing $I_\nu$ in the closed case.
Next we apply H\"older's inequality to
$\nabla F(\phi)= F'(\phi)\nabla\phi$, i.e.
$$
\| \nabla F(\phi)\|_1 \le \| F'(\phi)\|_{p',\ce}\|\nabla\phi\|_p.
$$
Finally equation~\ref{eq:rhoconcaveineq} tells us that the $\ce$ in the
norm of $F'(\phi)$ can be replaced by $\cv$ at a cost of introducing the
multiplicative factor $\rho_{\sup}^{1/p'}$.
\proofbox

\begin{corollary}\label{cr:sobolev} For any $f\in C^1_\Dirichlet(\cg)$
and any $\nu>p\ge 1$ we have
$$
\| \nabla \phi \|_p \ge c_{\nu,p} \| \phi \|_{p\nu/(\nu-p)}\qquad\mbox{where}
\quad c_{\nu,p} = I_\nu\rho_{\sup}^{-(p-1)/p}(\nu-p)/[p(\nu-1)].
$$
The same is true in the closed case if we add a tilde to $I_\nu$ and
further require $f$ to be split.
\end{corollary}
\proof We apply the above proposition with $F(x)=\ex{x}{r}$ where
$r=p(\nu -1)/(\nu-p)$.
\proofbox

\begin{theorem}  For $\phi\in C^1_\Dirichlet(\cg)$, we have for any $\nu>2$
$$
\| \nabla f \|_2 \ge (I_\nu \rho_{\sup}^{-1/2}/2) \| f \|_2^{1+(2/\nu)}
\| f \|_1^{-2/\nu}.
$$
The same is true in the closed case if we add a tilde to $I_\nu$ and
further require $\phi$ to be split.
\end{theorem}

\proof Applying proposition \ref{pr:general} to $\phi=f$ with $F(x)=\{x\}^2$ and $p=q=2$
gives
$$
2 \rho_{\sup}^{1/2} \|f\|_2 \| \nabla f\|_2 \ge I_\nu \|f\|^2_{2\nu'} .
$$
By H\"older's inequality we have $\| f^2\|_1\le\| f^\gamma\|_r
\|f^\delta\|_{r'}$ provided that $\gamma+\delta=2$.  Take $\gamma,\delta,
r,r'$ so that in addition we have\footnote{
i.e. take $\gamma=2(\nu'-2)/(\nu'-1)$, $\delta=2\nu'/(\nu'-1)$, 
and $r=(\nu'-1)/(\nu'-2)$.
}
$\gamma r=1$ and $\delta r'=2\nu'$, we have
$$
\|f\|_2^{2(2\nu'-1)/\nu'} \le \| f\|_1^{(2\nu'-2)/\nu'}\|f\|_{2\nu'}^2.
$$
Combining the two above displayed formulas yields the theorem.
\proofbox

We remark that in the above proof we could impose
$\delta r'=1$ and $\gamma r=2\nu/(\nu-2)$ and apply 
corollary~\ref{cr:sobolev} with $p=2$.  
This is often done in the literature, and yields a similar result but gives
the weaker constant of $c_{2,\nu}$ replacing $I_\nu \rho_{\sup}^{-1/2}/2$.

\begin{corollary}\label{cr:nashineq}
If $f\in C^1_\Dirichlet(\cg)$ with $\int f\,d\cv=0$ and $\cg$ closed,
then for any $\nu>2$ we have
$$
\| \nabla f \|_2 \ge  (\wt I_\nu \rho_{\sup}^{-1/2}/2) \| f \|_2^{1+(2/\nu)}
\| f \|_1^{-2/\nu} .
$$
\end{corollary}
\proof 
Let $a$ minimize $\|f-t\|_1$ as a function of t.
Let $\hat f=f-a$.  Then $\nabla f=\nabla \hat f$,
$\|\hat f\|_1\le \|f\|_1$, and $\|\hat f\|_2\ge \|f\|_2$, and the corollary
follows from applying the previous theorem to $\hat f$, since $\hat f$
is split.
\proofbox
The above theorem and corollary are
known as a {\em Nash inequality}, because it is
the main inequality necessary in Nash's method.  What is new here is
our simple proof of this inequality in the closed case, 
corollary~\ref{cr:nashineq}.

Corollary~\ref{cr:sobolev} is part of the Sobolev embedding theorem, and 
$c_{\nu,p}$ is called a Sobolev constant.  The optimal value of $c_{\nu,p}$
is quite interesting, and the value in the above corollary is certainly
not optimal for $\reals^n$ (and $\nu=n$) (see \cite{gilbarg-trudinger},
page 158).
However, we do know that the $c_{\nu,p}=0$ in the $\nu=p$ case, i.e.
there is no $\| \nabla \phi \|_p$ lower bound in terms of $\| \phi \|_\infty$
and $I_p$.
\begin{proposition}\label{pr:logbound}
Let $c^*_{\nu,p}$ be a constant depending only
on $\nu$ and $p$ such that 
corollary~\ref{cr:sobolev} holds for all $M,\phi$ with $c_{\nu,p}=
I_\nu\rho_{\sup}^{-1/p'}c^*_{\nu,p}$.  Then for any fixed $p$
there is a $C>0$ such that
$c^*_{\nu,p}\le C(\nu-p)^{(1/p)-1}$ for all $\nu$ sufficiently close to
and greater than $p$.
\end{proposition}
\proof We use a function $\phi$, growing logarithmically near a point,
just as in analysis.  The details are in appendix A.
\proofbox

Other common pieces of the Sobolev embedding theorem, in the analysis case, 
are the $\nu<p$ embedding theorems, stating
$\| \nabla \phi \|_p \ge c_{\nu,p} \| \phi \|_\infty$ for $c_{\nu,p}$
depending only on $p,\nu,I_\nu,\cv(\cg)$.
\begin{proposition}\label{pr:sob p>nu}
For any split function, $\phi$, and any $p>\nu\ge 1$ there is a constant
$c_{\nu,p}>0$ such that
$$
\| \nabla \phi \|_p \ge c_{\nu,p}\cv(\cg)^{(1/p)-(1/\nu)}I_\nu\rho_{\sup}^{-1/p'}
\| \phi \|_\infty
$$
No lower bound on $\| \nabla \phi \|_p$ is possible based on bounds on
$I_\nu$ and $\| \phi \|_\infty$ for arbitrary $\cg$ (i.e. $\cv(\cg)$
infinite).
\end{proposition}
\proof This is standard.  We string together a number of applications of
proposition~\ref{pr:general} to $F(x)=\{x\}^\gamma$ for various $\gamma$, as
in \cite{gilbarg-trudinger}.  The details are in appendix A.

Finally, one has the ``critical case'' of the Sobolev inequalities, namely
$p=\nu$, where, as we've said, $c_{\nu,\nu}=0$.  However, one can replace
this with various inequalities of exponential type, also known as Trudinger
inequalities (the first appears in \cite{trudinger}).  An example would be:
\begin{proposition} For any split function, $\phi$, we have that for any
$\gamma<1$
$$
\inte{\Bigl(\exp(\gamma\wt\phi)\Bigr)^{\nu'}} \le \cv(\cg) (1-\gamma)^{-\nu'} 
\quad\mbox{where}\quad
\wt\phi=\phi I_\nu \rho_{\sup}^{-1/\nu'}/\|\nabla\phi\|_\nu.
$$
\end{proposition}
\proof By proposition~\ref{pr:general} with $F(x)=\ex{x}{r}$ we have
$\|\wt\phi^r\|_{\nu'}\le r \|\wt\phi^{r-1}\|_{\nu'}$.  It follows that for
integers $r\ge 0$ we have $\|\wt\phi^r\|_{\nu'}\le r! \|1\|_{\nu'}$, and
so $\|\exp(\gamma\wt\phi)\|_{\nu'} \le \|1\|_{\nu'}/(1-\gamma)$;
the proposition easily follows.
\proofbox

\subsection{Nash's Method}
\label{subsecnash}

Now we use the method of Nash (see \cite{nash}) and its modification by
Cheng and Li (see \cite{cheng-li})
to obtain a diagonal heat kernel estimate.  

Now form 
$$
G(x,y,t) = \left\{ \begin{array}{ll} K(x,y,t) & \mbox{if non-closed,}\\
K(x,y,t)-1/\cv(\cg) & \mbox{if closed,} \end{array} \right.
$$
where $K$ is the minimal heat kernel.

\begin{definition} $H=H(x,y,t)$ defined for $x,y\in\cg$ and $t\ge 0$
is {\em heat-like} if (0) it is edgewise
linear
in $x,y$, (1) $H(x,x,t)\ge 0$,
(2) $H(x,y,t)=H(y,x,t)$, (3) $\Delta_x H(x,y,t)=\partial_t H(x,y,t)$ ,
and (4) $H$ is self-reproducing, i.e. 
$$
H(x,y,t)=\int H(x,z,t')H(z,y,t-t')\,\dv(z)
$$
for any $0<t'<t$, and we may partially differentiate this equation in $t$
and interchange the integral and differentiation.  
$H$ is {\em pre-Nash} if for any $y,t$ we have 
$f(x)=H(x,y,t)$ satisfies 
\begin{equation}\label{eq:needed_inequalities}
\| \nabla f \|_2 \ge C \| f \|_2^{1+(2/\nu)}\| f \|_1^{-2/\nu}
 \quad\mbox{and}\quad  \| f \|_1\le\gamma
\end{equation}
for some constants $C,\gamma>0$; if so then
$$
\| \nabla f \|_2 \ge C_1 \| f \|_2^{1+(2/\nu)} \quad\mbox{where}\quad 
C_1=C\gamma^{-2/\nu}.
$$
\end{definition}

Clearly $G$ is heat-like
\jpcom{this is not that clear : the trouble is that we can not use 
corollary 7.2 directly with $\phi(y) = K(x,y,t)$, since it does not belong
to $C_\Dirichlet(\cg)$. However there is an easy way to prove the theorem which follows by using the fact 
that the heat-kernel is the limit of a sequence of heat-kernels of finite graphs}
 in the non-closed case.  In the closed case
we may write:
$$
K(x,y,t) = \sum_i e^{-t\lambda_i}\phi_i(x)\phi_i(y)
$$
where $(\lambda_i,\phi_i)$ form a complete set of eigenpairs, and notice
that we may take $\lambda_1=0$.  Then $G$ becomes the above sum over
$i>0$, and it is easy to verify that $G$ is heat-like.

Since for fixed $y,t$ the integral of
$K(x,y,t)$ is $1$, we have that $G$ is pre-Nash with $C=I_\nu\rho_{\sup}^{-1/2}
/2$ and
$\gamma=1$ in the non-closed case 
\footnote{The left-hand inequality in
equation~\ref{eq:needed_inequalities} is clear if $\cg$ is finite.  In 
general, since $K$ is limit of finite heat kernels of graphs with 
isoperimetric constants no smaller than that of $\cg$, we may conclude
the same.}.
Since $G=K-\phi_1(x)\phi_1(y)$ in the
closed case, and $\phi_1(x)=\cv^{-1/2}(\cg)$, we have $G$ is pre-Nash
in the closed case with
$\gamma=2$ and $C=\wt I_\nu\rho_{\sup}^{-1/2}/2$.

\begin{theorem}[Nash] 
\label{th:nash} Let $G(x,y,t)$ be heat-like and pre-Nash.
Then we have 
$$
G(x,x,t)\le C_2 t^{-\nu/2}
\qquad\mbox{where}\qquad C_2=(\nu/2)^{\nu/2}C_1^{-\nu}
$$
\end{theorem}
\proof (See \cite{nash}.)  For any fixed $x\in M$ and $t>0$ we have
$$
G(x,x,t) = \| G(x,\cdot,t/2)\|_2^2 ,
$$
and so
$$
-(\partial/\partial t) G(x,x,t) = -\int G(x,y,t/2) G_t(x,y,t/2) \,\dv(y)
$$

$$
= \int G(x,y,t/2) \Delta_y G(x,y,t/2) \,\dv(y)
$$
 
$$
= \| \nabla_y G(x,y,t/2)\|_2^2 \ge C_1^2 
\| G(x,\cdot,t/2) \|_2^{2+(4/\nu)} .
$$
Furthermore,
$$
\| G(x,\cdot,t/2) \|_2^{2+(4/\nu)} = G(x,x,t)^{(2+\nu)/\nu}.
$$
Hence
$$
-G(x,x,t)^{-(2+\nu)/\nu}(\partial/\partial t) G(x,x,t)\ge C_1^2
$$
We conclude that
$$
G(x,x,t)^{-2/\nu} \ge G(x,x,0)^{-2/\nu} + (2/\nu)C_1^2 t
\ge (2/\nu)C_1^2 t.
$$
\proofbox

As a corollary we get an eigenvalue estimate in the closed case.  
Namely, we have
$$
ke^{-\lambda_kt} \le \sum_{i>0}e^{-\lambda_i t} = \int G(x,x,t)\,\dv(x)
\le C_2 t^{-\nu/2} \cv(\cg)
$$
for any $t>0$.  Taking $t=\nu/(2\lambda_k)$ yields
$$
\lambda_k\ge C_3 (k/\cv(\cg))^{2/\nu} \qquad\mbox{where}\qquad
C_3=C_2^{-2/\nu} \nu/(2e).
$$
We conclude:
\begin{corollary} In the closed case we have for any $\nu>2$
$$
\lambda_k \ge (k/\cv(\cg))^{2/\nu} C_2^{-2/\nu} \nu/(2e)
= (k/\cv(\cg))^{2/\nu} 2^{-4/\nu} (I_\nu\rho_{\sup}^{-1/2}/2)^2/e 
$$
\end{corollary}
We notice that the $\nu=\infty$ bound is weaker than Cheeger's inequality
by a factor of $e$, in that the above equation yields $\lambda_k\ge
(I_\infty/2)^2/e$.

We also get a heat kernel estimate in the non-closed case as well as the
closed case:
\begin{corollary} For any $\cg$ and any $\nu>2$ we have
$$
K(x,x,t)\le C_2t^{-\nu/2},
$$
where $C_2=(\nu/2)^{\nu/2}C_1^{-\nu}$ where $C_1=I_\nu\rho_{\sup}^{-1/2}/2$.
If $\cg$ is closed, the same holds with $K$ replaced by $K-1/\cv(\cg)$,
with a tilde added to $I_\nu$, and with $C_1$ multiplied by $2^{-2/\nu}$.
\end{corollary}
\proof The only new statement here is that the above inequality is also valid
when $\cg$ is infinite (i.e. the non-closed case).  But in this case we
simply note that the inequality holds for any $K_A(x,x,t)$ with $A$ a finite
subset of $\cg$'s vertices, and then take the limit as $A$ exhausts $\cg$.
\proofbox

\subsection{Other estimates}

The previous results gave us heat kernel and eigenvalue estimates, when one has an
isoperimetric inequality of the form $\area(\partial \Omega) \geq C \cv(\Omega)^{\alpha}$,
for some $0 < \alpha \leq 1$. We address now the issue of obtaining results of the same kind
when one has an isoperimetric inequality of the form 
\begin{equation}
\label{isoperimetric}
\area(\partial \Omega) \geq \frac{\cv(\Omega)}{\phi(\cv(\Omega))}
\end{equation}
where $\phi$ is a positive and non-decreasing function defined on $(0,\infty)$.

We will prove the following theorem (which will be a generalization of 
Theorem \ref{th:nash}).

\begin{theorem}
\label{th:general}
Let $$F(x) = \int_x^\infty \frac{\left( \phi (4/u) \right)^2}{u} du. $$ 
and $C=\frac{1}{32 \rho_{\sup}}$. 
If (\ref{isoperimetric}) holds for any admissible $\Omega$ , then
$$K(x,x,t) \leq F^{-1}(Ct).$$ 

\end{theorem}

\noindent
{\bf Remark} 
If we set $\phi(x)=x^{1/\nu}$, then we get Theorem \ref{th:nash} with 
slightly worse constants. 

The proof of this theorem is obtained through the following chain of implications,
where $C^1_\Dirichlet(\Omega) = \{f \in C^1_\Dirichlet(\cg) \; | \; f \equiv 0 \; \mbox{on} \; \complement{\Omega} \}$.

\begin{alignat}{2}
\forall  \; \mbox{admissible} \; \Omega, & \qquad
\area(\partial \Omega) & \geq & \frac{\cv(\Omega)}{\phi(\cv(\Omega))} \label{eq:iso}\\
 & \qquad \Downarrow & &  \nonumber \\
\forall f \in C^1_\Dirichlet(\Omega), & \qquad 
\phi(\cv (\Omega)) \| \nabla f \|_{1,\ce} & \geq & \|f\|_{1,\cv} \label{eq:ff}\\
 & \qquad \Downarrow & & \nonumber \\
\forall f \in C^1_\Dirichlet(\Omega), & \qquad 
2 \rho_{\sup}^{1/2} \phi(\cv (\Omega)) \| \nabla f \|_{2,\ce} & \geq &\|f\|_{2,\cv} \label{eq:fk}\\
 & \qquad \Downarrow & &  \nonumber \\
\forall f \in C^1_\Dirichlet(\cg)  ), & \qquad
32 \rho_{\sup}\left(  \phi\left(\frac{4 \|f\|_{1,\cv}^2}{\|f\|_{2,\cv}^2} \right) \right)^2
\| \nabla f \|_{2,\ce}^2  & \geq &\|f\|_{2,\cv}^2 
\label{eq:gennash}\\
 & \qquad \Downarrow & & \nonumber \\
\forall x \in V, \;t > 0  & \qquad K(x,x,t) & \leq & F^{-1}(Ct)\label{eq:esti}
\end{alignat}

\proof {\bf (\ref{eq:iso}) $\Rightarrow$ (\ref{eq:ff})}
This follows since $\phi$ is increasing, so we have $I_\infty(\Omega)=
1/\phi(\cv(\Omega))$.

\jpcom{this is done by using a slightly different version of the coarea formula}

\proof{\bf (\ref{eq:ff}) $\Rightarrow$ (\ref{eq:fk})}
This proof is essentially identical to that for proposition~\ref{pr:general}, and is done by applying 
(\ref{eq:ff}) to $f^2$, then using Cauchy-Schwartz, and corollary
{eq:rhoconcaveineq}.

\jpcom{this is almost proposition \ref{pr:general}, and can be proved in a similar fashion}

\proof{\bf (\ref{eq:fk}) $\Rightarrow$ (\ref{eq:gennash})}
This is basically proposition 10.3 of \cite{BCLS} in our context. The proof goes as follows 
(we will omit the subscript $\cv$ in  expressions like $\|f\|_{\alpha,\cv}$ ):
first we notice that for $t>0$ 
\begin{eqnarray*}
\|f\|_2^2 & = & \int_{|f| \geq 2t} f^2 d\cv + \int_{|f| \leq 2t} f^2  d\cv \\
&  \leq & 
\int_{|f| \geq 2t} 4 \left((|f|-t)^+\right)^2 d\cv +  \int_{|f| \leq 2t} f^2 d\cv\\
&  \leq & 
4 \|(|f|-t)^+\|_2^2 +  \int_{|f| \leq 2t} f^2 d\cv\\
&  \leq & 
4 \|(|f|-t)^+\|_2^2 +  2t \|f\|_1 
\end{eqnarray*}
We use \jpcom{this has to be justified with some care} (\ref{eq:fk}) with
$(|f|-t)^+$ and we obtain\footnote{
By approximation we have that equation~\ref{eq:fk} implies the same inequality
for any Lipschitz function vanishing out of $\Omega$.  Hence we can apply
this inequality to $(|f|-t)^+$.
}
\begin{eqnarray*}
\|(|f|-t)^+\|_2^2 
& \leq & 
4 \rho_{\sup} \left(\phi(\cv(|f|\geq t))\right)^2 \|\nabla (|f|-t)^+ \|_2^2 \\
& \leq &
4 \rho_{\sup} \left(\phi(\cv(|f|\geq t))\right)^2 \|\nabla f \|_2^2 \\
& \leq & 
4 \rho_{\sup} \left(\phi(\frac{\|f\|_1}{t})\right)^2 \|\nabla f \|_2^2
\end{eqnarray*}
Therefore
$$ \|f\|_2^2 \leq 
16 \rho_{\sup} \left(\phi(\frac{\|f\|_1}{t})\right)^2 \|\nabla f \|_2^2 + 2t \|f\|_1.$$
We choose $t$ to be $\frac{\|f\|_2^2}{4 \|f\|_1}$
to conclude the proof.

\proof{\bf (\ref{eq:gennash}) $\Rightarrow$ (\ref{eq:esti})}
\jpcom{This is just the sketch of the proof : we need the fact that the heat-kernel is a limit
of a sequence of heat-kernels which are defined on finite graphs.}
We may assume $\cg$ is connected.  First assume that $\cg$ is finite.
By assumption, $\partial\cg$ is non-empty (or equation~\ref{isoperimetric} is
immediately violated).
Let $U(t)=K(x,x,t)$. As in the proof of theorem \ref{th:nash} 
we note that 
$$ - \frac{dU}{dt} = \| \nabla_y K(x,y,t/2)\|_2^2.$$
Let $f(y) = K(x,y,t/2)$.
By first using the non-negativity of the heat kernel, and second its
$L^{\infty}$contractivity we obtain
$$\|f\|_1 = \int |K(x,y,t/2)| d\cv(y) = \int K(x,y,t/2) d\cv(y) \leq 1$$
Moreover 
$$\|f\|_2^2 = \int K(x,y,t/2)^2 d\cv(y) = K(x,x,t)=U(t).$$
We now apply (\ref{eq:gennash}) to $f$ 
\jpcom{it's exactly here that the graph better be finite...}
$$32 \rho_{\sup} \left(  \phi\left(\frac{4 \|f\|_1^2}{\|f\|_2^2} \right) \right)^2
\| \nabla f \|_2^2   \geq \|f\|_2^2 $$
and obtain by using the previous remarks 
$$ -\frac{dU}{dt} (\phi(4/U))^2 \geq C U.$$
Dividing both sides of this inequality by $U$ and integrating  against $dt$ from $0$ to $t$ we get
$$\int_{U(t)}^{U(0)} \frac{(\phi(4/U))^2}{U} dU \geq \int_0^t C dt$$
Therefore 
$$ F(U(t)) = \int_{U(t)}^{\infty} \frac{\phi(4/U))^2}{U} \geq Ct,$$
we conclude the theorem for finite $\cg$
by using the fact that $F$ is decreasing.

If $\cg$ is infinite, we see that the bound applies to $K_A(x,x,t)$ for 
finite $A\subset\interior V$ (since clearly equation~\ref{isoperimetric}
also holds for all admissible $\Omega$ in $\cg_A$), with notation as in
subsection~\ref{sb:finite_limit}.  We conclude the 
theorem by taking the limit as $A$ exhausts $\cg$.
\proofbox

\appendix
\section{Some Sobolev Related Calculations}

\subsection{Logarithmic functions on radial graphs.}

In this section we use $f\approx g$ to mean there exist universal constants
$c_1,c_2>0$ such that $c_1f\le g\le c_2 f$.  We use $f\asle g$ to mean there
is a universal constant $c_1>0$ such that $f\le c_1 g$, and similarly for
$f\asge g$.

\subsubsection{A non-closed example}
Here we give a sequence of non-closed graphs (in fact finite but with
boundary) which bound the optimal constant in proposition~\ref{pr:logbound}.
\begin{definition} The {\em path of length $n$} is
the graph $G=(V,E)$ with $V=\{1,\ldots,n\}$,
and an edge $\{i,i+1\}$ for each $i=1,\ldots,n-1$.  
For $\nu\ge 1$, by the {\em $\nu$-dimensional radial graph of size $n$}, 
$G=G_{n,\nu}$, we understand
the path of length $n$ with $n$ being a boundary vertex,
with the following measures, $\cv$ and
$\ce$: $\ce(\{i,i+1\})=i^{\nu-1}$, and $\cv$ is taken {\em natural}
with respect
to $\ce$, meaning we take the unique measure $\cv$ such that $\cg$ is
$2$-regular.
\end{definition}
In other words, $\cv(i)=[(i-1)^\nu + i^\nu]/2$ for $i<n$ and
$\cv(n)=(n-1)^\nu/2$.
We can extend the definition above to the $n=\infty$ case, giving the
(countably) infinite path and the 
(countably) infinite $\nu$-dimensional radial graph in the obvious way.

First we observe:
\begin{proposition} For $1\le \nu\le n$ (allowing $n=\infty$) we have
$I_\nu(G)\approx \nu^{1/\nu'}$ for $G=G_{n,\nu}$.
\end{proposition}
\proof
According to Yau's remark, it suffices to consider $\Omega$ connected, i.e.
$\Omega=(a,b)$  with $1 < a < b < n$, or $\Omega=[1,b)$.  
Let $i$ and $j$ be respectively the smallest integer and the biggest integer, such that
$[i,j] \subset \Omega$.
Note that 
$$
 \area(\partial \Omega) = (1/2)\Bigl( (i-1)^{\nu-1}+j^{\nu-1} \Bigr) 
\approx j^{\nu-1},
$$
and
$$
\cv (\Omega) = \sum_{t=i}^j [(t-1)^{\nu-1} + t^{\nu-1}]/2 \approx \nu j^\nu,
$$
and hence
$$
i_\nu(\Omega)\approx \nu^{1/\nu'}.
$$

\ignore{
If $j=n$ we have
$$
|\partial \Omega|= (i-1)^{\nu-1}/2.
$$
But then $|\wb\Omega|\ge |G|/2$ implies $i^\nu \asge n^\nu/2$ so
$i\ge n c^{1/\nu}$ so
$$
(i-1)^{\nu-1} \ge \Bigl(n-c^{-1/\nu}\Bigr)^{\nu-1} c^{(\nu-1)/\nu}
$$
for some absolute constant, $c$.  It follows that $(i-1)^{\nu-1}\asge 
n^{\nu-1}$; since $(i-1)^{\nu-1}<n^{\nu-1}$ we have $|\partial\Omega|\approx
j^{\nu-1}$.  The rest of the calculation proceeds as before.

We conclude that $i_\nu(\Omega)\approx \nu^{1/\nu'}$ for all $\Omega$ as
above, and so $I_\nu\approx \nu^{1/\nu'}$.
}
\proofbox

Define $f_m$ on $G$ as above for any $m\le n$ via
$$
f_m(i) = \left\{\begin{array}{ll} \log(m/i) &\mbox{for $i\le m$,} \\
		0 &\mbox{for $i>m$.} \end{array} \right.
$$
We have
$$
\|\nabla f_m\|_p^p = \sum_{i=1}^{m-1} \log^p\Bigl(1+(i+1)^{-1}\Bigr) i^{\nu-1} 
\approx \sum_{i=1}^{m-1} i^{\nu-p-1} \;\approx\; \left\{\begin{array}{ll}
m^{\nu-p}/(\nu-p) &\mbox{if $\nu>p$,} \\
\log m & \mbox{if $\nu=p$,} \\
1 & \mbox{if $\nu<p$.}\end{array}
\right.
$$
Also
$$
\|f_m\|_q^q = \sum_{i=1}^{m-1} \Bigl(\log(m/i)\Bigr)^q i^{\nu-1} \asge
\Bigl(\log(m/k)\Bigr)^q k^\nu/\nu
$$
for any $k=1,\ldots m$.  Taking $k=me^{-q/\nu}$ (assuming $me^{-q/\nu}\ge 1$)
yields:
$$
\|f_m\|_q \asge (q/\nu) m^{\nu/q} \nu^{-1/q}.
$$
Of course $\|f_m\|_\infty=\log m$.  

Taking $q=p\nu/(\nu-p)$ for $\nu\ge p$
(meaning $q=\infty$ with $\nu=p$), 
yields the first part of proposition~\ref{pr:logbound}.
Taking $q=\infty$ for $\nu<p$ yields the last claim of 
proposition~\ref{pr:sob p>nu}.

\subsubsection{A Closed Example}
In this section we give a sequence of closed graphs demonstrating the same
bound on the optimal constant in proposition~\ref{pr:logbound}.  We do so
by the process of {\em doubling}.
\begin{definition} Let $G$ be a graph.  By the {\em double} of $G$ we mean
the graph $H$ formed from taking two copies of $G$, $G_+,G_-$, (which we
sometimes call the positive and negative parts) and by
identifying the boundary points of $G_+$ with those of $G_-$ and declaring
such points to be no longer boundary points.  $H$ comes with a natural
involution $\iota$ exchanging $G_+$ and $G_-$.  The $\cv$ measure on the
boundary of $G$ is doubled in $H$ to preserve naturality.
\end{definition}

\begin{proposition} Let $f$ be an odd function on a doubled graph, i.e.
$f(\iota(x))=-f(x)$ for all $x$.  Then $f$ is split and $p$-balanced for
all $p$.
\end{proposition}
\proof Clearly $\intv{\ex{f}{p-1}}=0$, and all claims follow.
\proofbox

\begin{definition} The {\em $n$-th doubled $\nu$-dimensional radial graph},
$\wt G_{n,\nu}$ is the double of $G_{n,\nu}$.
\end{definition}
We similarly show that $I_\nu(\wt G_{n,\nu})\approx \nu^{1/\nu'}$
(it suffices to take connected $\Omega$ with $|\Omega|\le |\wt G_{n,
\nu}|/2$).
Now we take $\wt f_m$ on $\wt G_{n,\nu}$ to be $f_m$ on the positive
part and $-f_m$ on the negative part.  Clearly $\wt f_m$ is odd, and
therefore split and $p$-balanced for all $p$.  The calculations of
$\|\wt f_m\|_q$ and $\|\nabla \wt f_m\|_p$ go through as for $f_m$.

\subsection{Classical graphs and the $\nu$-dimensional graph}

In the previous subsection we found graphs demonstrating the
best possible constant in proposition~\ref{pr:logbound}.  But these
graphs weren't classical graphs, in that their measures weren't
traditional.  We can modify these graphs to obtain traditional regular
measures with the same results (e.g. as in proposition~\ref{pr:logbound}).

Let $V(1),\ldots,V(n)$ be a non-decreasing sequence
of positive rationals such that $V(n)=2[V(n-1)-V(n-2)+\cdots]$.
Let $G$ be a classical graph as follows: its
vertices are partitioned into $n$ sets,
$V_1,\ldots,V_n$, with $|V_i|=k V(i)$ for an integer $k$ making $k V(i)$
integral for all $i$; its edges are
$E(i)$ copies of a complete graph between $V_i$ and $V_{i+1}$ for each $i$, 
where
$E(i)=\ell [k V(i)V(i+1)]^{-1} [V(i)-V(i-1)+V(i-2)-\cdots]$, for an $\ell$
which makes the $E(i)$ integers.
Now declare $V_n$ to be the boundary points of $G$, and form the double,
$H$.  Then $H$ is an $\ell$-regular graph.  Clearly for any $\nu$ we have
$I_\nu$ is the same for $H$ as it is for the double, $\wt H$,
of the path of length
$n$ (with boundary point $n$) and which is $\ell$-regular and with
measure $\cv(i)=k V(i)$.  And clearly a function, $f$, on $\wt H$ lifts
to a function on $H$ which preserves $\|f\|_q$ and $\|\nabla f\|_p$
for any $p$ and $q$.

In the above paragraph, fix a $\nu$ and an integer $m$, and set
$V(i)=\lfloor m i^{\nu-1} \rfloor$ for $1\le i\le n-1$, and
$$
V(n)=2[V(n-1)-V(n-2)+\cdots].
$$
Then we see $V(n)=mn^{\nu-1}[1+o(1)]$, and it follows that the calculations
of the previous subsection for the optimal constant in 
proposition~\ref{pr:logbound} all hold here.

\subsection{The $p<\nu$ Sobolev Imbedding}

Here we prove proposition~\ref{pr:sob p>nu}.  By scaling $\cv$ and $\ce$ we
may assume $\cv(\cg)=1$ and  $\rho_{\sup}=1$.
As in \cite{gilbarg-trudinger}
page 156,
set $\wt u = I_\nu u / \|\nabla u\|_p$.  Since $|M|=1$ we have
$\|\nabla u\|_1\le \|\nabla u\|_p$, and so $\|\wt u\|_{\nu'}\le 1$.  
By proposition~\ref{pr:general}
$$
\|u^\gamma\|_{\nu'} I_\nu \le \gamma\|u^{\gamma-1}\|_{p'}\|\nabla u\|_p,
$$
i.e.
$$
\| \wt u^\gamma\|_{\nu'} \le \gamma \|\wt u^{\gamma-1}\|_{p'},
$$
i.e.
\begin{equation}\label{eq:sobiterate}
\|\wt u\|_{\gamma\nu'} \le \gamma^{1/\gamma} \|\wt u\|_{p'(\gamma-1)}^{1-(
1/\gamma)},
\end{equation}
provided that $p'(\gamma-1)\ge 1$ (this is the convexity of $(F')^{p'}$).
Set $\gamma_i=1+\delta+\cdots+\delta^{i+1}$ where $\delta=\nu'/p'>1$;
we have $p'(\gamma_0-1)=\nu'$ and $p'(\gamma_i-1)=\nu'\gamma_{i-1}$.
Applying equation~\ref{eq:sobiterate} $n$ times we have for any $n$
$$
\|\wt u\|_{\gamma_n\nu'} \le \Bigl(\gamma_0 \gamma_1^{1/\delta}\cdots 
\gamma_n^{1/\delta^n} \Bigr)^{\delta^n/\gamma^n}
$$
using the fact that $\|\wt u\|_{\nu'}\le 1$.
It follows that for any $n$
$$
\|\wt u\|_{\gamma_n\nu'} \le c_1^{c_2}
$$
where 
$$
c_1=\gamma_0 \gamma_1^{1/\delta}\gamma_2^{1/\delta^2}\cdots  \quad\mbox{and}
\quad c_2= p'(\nu'-p')/(\nu')^2.
$$
Taking $n\to\infty$ yields the proposition.
\proofbox

\section{Proofs of Uniqueness and Non-uniqueness}

Notice that if $u_1,u_2$ are two solutions to the Dirichlet initial value
problem with same initial value, then $u=u_1-u_2$ satisfies the same with
initial value $0$.  So to prove uniqueness it suffices to prove uniqueness
in this case.  Similarly for non-uniqueness.

We begin by proving theorem~\ref{th:finite_uniqueness}.  If $u(x,t)$
solves the Dirichlet initial value problem in $[0,T]$ with initial value
$0$, then assuming $u>0$ somewhere, we have that $u$'s maximum value, $M$,
over
$\cg\times [0,T]$ is attained somewhere (since $\cg$ is finite), 
say $u(x_0,t_0)=M$ (of course, $t_0>0$ and $x\in\interior V$).
Arguing as in theorem~\ref{th:simple_maximum_principle} we have
$u_t(x_0,t_0)\ge 0$ and so $\Delta u(x_0,t_0)\le 0$ and we conclude
$u(w,t_0)=M$ for all $w\in\interior V$ with an edge to $x_0$.  Similarly
we conclude $u(w,t_0)=M$ for all $w\in\interior V$ with a path to $x_0$
in $\interior V$.

Consider that
$$
\frac{d}{dt}\int u^2(x,t)\,d\cv(x)=\int 2u(x,t)u_t(x,t)\,d\cv(x)
$$
$$
= - \int 2u(x,t)\Delta u(x,t)\,d\cv(x) =  - \int |\nabla u|^2\,d\cv(x)\le 0.
$$
In brief, $\int u^2\,d\cv(x)$ is non-increasing.  In the above we could have
assumed that $t_0$ was as small as possible, i.e. that $u(x,t)<M$ for $t<t_0$
(again using the fact that $\cg$ is finite).
Since $u$ is differentiable in $t$ and $\cg$ is finite, we have that
$u(x,t)$ remains positive for all $x$ and $t$ near $t_0$.
But then $\int u^2(x,t)\,d\cv(x)$ is visibly less for $t$ slighly less
than $t_0$ than it is for $t=t_0$, a contradiction.
\proofbox

Next we prove theorem~\ref{th:mild_graph_nonuniqueness}.
We proceed as in \cite{friedman-parabolic}.
It is known (see \cite{mandelbrojt}) that for any $\delta>0$ there is a
non-zero
function $f\in C^\infty(\reals)$ vanishing outside of $(0,1)$ and with
$|f^{(m)}(t)|\le C^mm^{(1+\delta)m}$ for a constant $C$.  Let $\cg$ be
the graph whose vertices are the integers, $\integers$,
and with an edge $\{i,i+1\}$ for each $i\in\integers$; endow $\cg$ with
the standard $\cv$, $a_e$, and $\ell_e$, and take the boundary to be
empty.  Then for $x\in\integers$,
$$
u(x,t) = \sum_{m=0}^\infty f^{(m)}(t) \binom{x+m}{2m}
$$
is a finite sum; extending to $x\notin\integers$
by linearity, it is easy to verify that
$u$ satisfies the heat equation.  Since $u(x,0)=0$, this means the 
Dirichlet initial value problem does not have a unique solution.
(We easily see that
$|u(x,t)|\le (C_1|x|)^{|x|(1+\delta)+1}$.)
\proofbox

Next we prove theorem~\ref{th:growing_degree_uniqueness}
Let $u(x,t)$ be a solution to the Dirichlet initial value problem with 
initial value $0$ whose absolute value
is bounded in $[0,T]$ by a finite constant $B$.

We claim, by induction, that for any integer $j\ge 0$ we have
\begin{equation}\label{eq:iterated_bound}
|u(x,t)|\le BL_0(v)\cdots L_{j-1}(v)(2t)^j/j!
\end{equation}
For $j=0$ this just says $|u|\le B$, which is our assumption.  Now assume
the assumption holds for some $j$.  Write
$$
|\Delta u(x,t)| \le L(v)(|u(x,t)|+\sup_{y\sim x}|u(y,t)|)\le 
$$
$$
L(x)2B\sup_{y\sim x}L_0(y)\cdots L_{j-1}(y)(2t)^j/j!\le
2B L_0(x)\cdots L_j(x)(2t)^j/j!,
$$
and conclude
$$
|u(x,t)|=\left| \int_0^t u_s(x,s)\,ds \right|\le 
2^{j+1}B L_0(x)\cdots L_j(x)\int_0^t s^j/j!\,ds.
$$
We conclude equation~\ref{eq:iterated_bound} holds for $j+1$.

Now we take $j\to\infty$.  We conclude that there is an $\epsilon>0$ 
(depending only
on the $C$ in the theorem's hypothesis on $L$) such that $u(x,t)=0$
for all $x$ and $t\le\epsilon$.  But then we again conclude $u(x,t)=0$
for $t$ less than any multiple of epsilon, i.e. for all $t$.
\proofbox

Next we prove theorem~\ref{th:growing_degree_nonuniqueness}.  Let
$n_i=\lfloor i^{1+\alpha}\rfloor$ for $i$ a positive integer, so that
$n_i$ is a positive integer with $n_i$ growing like $i^{1+\alpha}$.
Let
$\cg$ be the the tree whose vertices are as follows: it has one vertex
labelled ``1.''  The vertex ``1'' has $n_1$ edges to $n_1$ different
vertices, each labelled ``2.''  Each vertex labelled ``2'' has $n_2$ edges
to $n_2$ distinct vertices labelled ``3,'' for a total of $n_1n_2$ vertices
labelled ``3.''  Similarly there are $n_1\cdots n_i$ vertices labelled
``$i+1$,'' each with one edge to an ``$i$'' vertex, and with $n_{i+1}$
edges to ``$i+2$'' vertices.

Consider a function $f$, on $\cg$, whose value at the ``$i$'' vertices
is $f(i)$.  We wish to make $f$ satisfy $\Delta f = -f$ at all vertices
which for ``$i$'' vertices with $i>1$ entails:
$$
[f(i)-f(i-1)] + [f(i)-f(i+1)]n_i = -f(i),
$$
or
$$
f(i+1) = [(2+n_i)f(i)-f(i-1)]/n_i,
$$
and which similarly entails $f(2)=(1+n_1)f(1)/n_1$ at the ``1'' vertex.
So choose $f(1)=1$, let $f(2)=(1+n_1)f(1)/n_1$
and let $f(i)$ for $i\ge 2$ be determined
recursively by the above equation.  We easily see by induction
that $f(i)<f(i+1)$.  Also clearly
$$
f(i+1) \le (2+n_i)f(i)/n_i,
$$
and so
$$
f(i) \le (1+2n^{-1}_{i-1})\cdots(1+2n^{-1}_{1}).
$$
Since $\alpha>0$ we have
$$
(1+2n^{-1}_{1})(1+2n^{-1}_{2})\cdots < \infty,
$$
and so the $f(i)$ are bounded.

It follows that $u(i,t)=e^tf(i)$ satisfies the heat equation, and the
$L^\infty$ norm of $u$ increases as a function of $t$.  
But by
our heat kernel construction we know
that the Dirichlet initial value problem with initial condition $f$ has
a solution $w(i,t)$ whose $L^\infty$ norm is no more than that of $f$.
So any multiple of $u-w$ is a non-zero solution of the heat equation with
zero boundary conditions and which is bounded for any fixed $t$.


\end{document}